# Linking Economic Complexity, Institutions and Income Inequality

D. Hartmann[1,2,3]*, M.R. Guevara[1,4,5], C. Jara-Figueroa[1], M. Aristarán[1], C.A. Hidalgo[1]*†

Draft, September 2016

Affiliations:
[1] Macro Connections, The MIT Media Lab, US.
[2] Fraunhofer Center for International Management and Knowledge Economy, Leipzig, DE.
[3] Chair for Innovation Management und Economics, University of Leipzig, DE.
[4] Department of Computer Science, Universidad de Playa Ancha, Chile.
[5] Department of Informatics, Universidad Técnica Federico Santa María, Chile.

*Correspondence to: hartmado@mit.edu, hidalgo@mit.edu
†current address: The MIT Media Lab, 75 Amherst Street, Cambridge, MA 02139.

ABSTRACT

A country's mix of products predicts its subsequent pattern of diversification and economic growth. But does this product mix also predict income inequality? Here we combine methods from econometrics, network science, and economic complexity to show that countries exporting complex products—as measured by the Economic Complexity Index—have lower levels of income inequality than countries exporting simpler products. Using multivariate regression analysis, we show that economic complexity is a significant and negative predictor of income inequality and that this relationship is robust to controlling for aggregate measures of income, institutions, export concentration, and human capital. Moreover, we introduce a measure that associates a product to a level of income inequality equal to the average GINI of the countries exporting that product (weighted by the share the product represents in that country's export basket). We use this measure together with the network of related products—or product space—to illustrate how the development of new products is associated with changes in income inequality. These findings show that economic complexity captures information about an economy's level of development that is relevant to the ways an economy generates and distributes its income. Moreover, these findings suggest that a country's productive structure may limit its range of income inequality. Finally, we make our results available through an online resource that allows for its users to visualize the structural transformation of over 150 countries and their associated changes in income inequality between 1963 and 2008.

# 1. INTRODUCTION

Is a country's ability to both generate and distribute income determined by its productive structure? Economic development pioneers, like Paul Rosenstein-Rodan, Hans Singer, and Albert Hirschman, would have said yes, since they argued in favor of a connection between a country's productive structure, and its ability to generate and distribute income. These pioneers emphasized the economic role of "structural transformations"—the process by which economies diversify from agriculture and extractive industries to more sophisticated forms of services and manufacturing (Rosenstein-Rodan, 1943; Singer, 1950; Hirschman, 1958).

But testing the intuition of these development pioneers has not been easy due to the complexity of measuring a country's productive structure. During the twentieth century, scholars did not go beyond simple quantitative approaches, such as (a) measuring the fraction of an economy employed in agriculture, manufacturing, or services; (b) using aggregate measures of diversity and concentration (Hirschman, 1945; Herfindahl, 1950; Imbs & Wacziarg, 2003); or (c) looking at diversification into related and unrelated varieties—that is, diversification into similar or different products (Teece et al., 1994; Frenken, Oort & Verburg, 2007; Saviotti & Frenken, 2008; Boschma & Iammarino, 2009). These measures of a country's productive structure, however, fail to take the sophistication of the products into account, or capture differences in industrial structures in a manner that is too coarse (i.e. by defining broad categories such as agriculture, manufacturing, and services).

Recently, though, the introduction of measures of 'economic complexity'—which we define and explain in the data and methods section below—have expanded our ability to quantify a country's productive structure and have revived interest in the macroeconomic role of structural transformations (Rodrik, 2006; Hausmann, Hwang & Rodrik, 2006; Hidalgo et al., 2007; Hidalgo & Hausmann, 2009; Felipe, 2009; Abdon & Felipe, 2011; Bustos et al., 2012; Caldarelli et al., 2012; Tacchella et al., 2012; Cristelli et al., 2013; Hausmann et al., 2014; Cristelli, Tacchella & Pietronero, 2015). These measures of economic complexity have received wide attention because they are highly predictive of future economic growth (ibid.). This also makes these measures of economic complexity relevant for social welfare, since economic growth and average income are correlated with country's absolute levels of poverty and social welfare (Bourguignon, 2004; Ravallion, 2004).

However, there are also multiple reasons why the productive structures of countries could be associated not only with economic growth, but also with a country's average level of income inequality.

First, the mix of products that an economy makes constrains the occupational choices, learning opportunities, and bargaining power of its workers and unions. Notably, in several emerging economies, technological catch-up and industrialization have provided new jobs and learning opportunities for workers, contributing to the rise of a new middle class (Milanovic, 2012). Conversely in several "industrialized" economies, de-industrialization, de-unionization, and rising global competition for the export of industrial goods have contributed to higher levels of income inequality. In the industrialized economies many industrial workers have become unemployed or were forced to work at low paying jobs, and the ability of unions to compress wage inequality has decreased (Gustafsson & Johansson, 1999; Acemoglu, Aghion & Violante, 2001).

Second, recent work on productive structures has highlighted that the complexity and diversity of products a country exports are a good proxy of the knowledge and knowhow available in an economy that is not captured by aggregate measures of human capital (Hidalgo, 2015)—such as the years of schooling or the percentage of the population with tertiary education. Moreover, productive structures can also be understood as a proxy of an economy's level of social capital and the health of its institutions, since the ability of a country to produce sophisticated products also critically depends on the ability of people to form social and professional networks (Hidalgo, 2015, Fukuyama 1995). For this reason, complex industrial products also tend to require a large degree of tacit knowledge and more distributed knowledge than found with simple products that are mainly based on resource richness or low labor costs. More distributed knowledge and a large degree of tacit knowledge can enhance the incentives to unionize and increase the effectiveness in negotiating high wages and therefore compress wage inequality.

Third, in a world in which economic power begets political power, non-diverse economies—such as countries with incomes largely based on few natural resources—are more susceptible to suffer from both economic and political capture (Engerman & Sokoloff, 1997; Collier, 2007; Hartmann, 2014).

Here, we contribute to the literature on economic complexity, income inequality, and structural transformations, by documenting a strong, robust, and stable correlation between a country's level of economic complexity (as proxied by the Economic Complexity Index) and its level of income inequality between 1963 and 2008. We find this correlation is robust to controlling for a variety of factors that are expected to explain cross-country variations in income inequality, such as a country's level of education, institutions, and export concentration. Moreover we find that, over time, countries that experience increases in economic complexity are more likely to experience decreases in their level of income inequality. We develop a product level index to estimate the changes in the level of income inequality that we would expect if a country were to modify its product mix by adding or removing a product. Our results suggest that a country's level of income inequality may be conditioned by its productive structure.

The remainder of the paper is structured as follows. Section 2 reviews the literature on economic development, institutions and income inequality. Section 3 presents the data and methods used in this paper. Section 4 compares the correlations between Gini and a variety of measures of productive structures, including the Economic Complexity Index (Hidalgo & Hausmann, 2009), the Herfindahl-Hirschman Index (Hirschman, 1945, Herfindahl, 1950), Entropy (Shannon, 1948), and the Fitness Index (Tacchella et al., 2012). This section then uses multivariate regressions and panel regressions to estimate the correlation between economic complexity and income inequality that is not explained by the correlation between income inequality and average income, population, human capital (measured by average years of schooling), export concentration, and formal institutions. Finally, Section 5 introduces an estimator of the level of income inequality expected for the exporters of 775 different products in the Standard Industrial Trade Classification at the four-digit level (SITC-4 Rev.2). We use this estimator to illustrate how changes in a country's productive structure are associated with changes in income inequality. Section 6 provides concluding remarks.

## 2. CONNECTING INCOME INEQUALITY AND ECONOMIC DEVELOPMENT

Decades ago Simon Kuznets (1955) proposed an inverted-u-shaped relationship describing the connection between a country's average level of income and its level of income inequality. *Kuznets' curve* suggested that as an economy develops, market forces would first increase and then decrease income inequality. Yet, Kuznets' curve has been difficult to verify. The inverted-u-shaped relationship predicted by Kuznets fails to hold if several Latin American countries are removed from the sample (Deininger & Squire, 1998), and in recent decades, the upward side of Kuznets' curve has vanished as inequality in many low-income countries has increased (Palma, 2011). Moreover, several East-Asian economies have grown from low to middle incomes while reducing income inequality (Stiglitz, 1996). Together, these findings undermine the empirical robustness of Kuznets' curve, and reaffirm that GDP per capita is an insufficient measure of economic development in terms of explaining variations in income inequality (Kuznets, 1934; Kuznets, 1973; Leontieff, 1951; Stiglitz, Sen & Fitoussi, 2009).

The empirical failure of Kuznets' curve resonates with recent work arguing that inequality is not only dependent on a country's rate or stage of growth, but also on its type of growth and institutions (Engerman & Sokoloff; 1997; Fields; 2002; Bourguignon, 2004; Ravallion, 2004; Sachs, 2005; Beinhocker, 2006; Collier 2007; Stiglitz, Sen & Fitoussi 2009; Acemoglu & Robinson, 2012; Hartmann, 2014). We should expect, then, that more nuanced measures of economic development (such as those focused on the sophistication of the products that a country exports) should provide information on the connection between economic development and income inequality that exceeds the limitations of aggregate output measures like GDP.

Understanding the determinants of income inequality is not simple since income inequality depends on a variety of factors, from an economy's factor endowments, geography, institutions and social capital, to its historical trajectories, changes in technology, and returns to capital (Engerman & Sokoloff; 1997; Gustafsson & Johansson, 1999; Acemoglu, Aghion & Violante, 2001; Fields, 2002; Beinhocker, 2006; Collier 2007; Davis, 2009; Acemoglu & Robinson, 2012; Brynjolfsson & Afee, 2012; Stiglitz, 2013; Frey & Osborne, 2013; Piketty, 2014; Autor, 2014, Hartmann, 2014).

Measuring these factors directly is difficult, but we can create indirect measures of them by leveraging the fact that the presence of these factors is expressed in a country's mix of products (Innis, 1970; Engerman & Sokoloff, 1997; Hausmann & Rodrik, 2003; Rodrik, 2006; Hausmann, Hwang & Rodrik, 2006; Hidalgo et al., 2007; Hidalgo & Hausmann, 2009; Felipe et al., 2012; Tacchella et al., 2012; Cristelli et al., 2013; Hausmann et al., 2014; Hidalgo, 2015). For example, post-colonial economies specializing in a limited number of agricultural or mineral products, like sugar, gold, and coffee, tend to have more unequal distributions of political power, human capital, and wealth (Innis, 1970; Engerman & Sokoloff, 1997; Acemoglu & Robinson, 2012), and hence, their productive structures provide us with indirect information about their geographies, human capital, and institutions. Conversely, sophisticated products, like medical imaging devices or electronic components, are typically produced in diversified economies with inclusive institutions and high levels of human capital. This means that the presence of complex industries in an economy, in addition to indicating the inclusiveness of that economy's institutions, also reveals the knowledge and knowhow that is embodied in its population (Hidalgo, 2015).

The idea that productive structures co-evolve with the inclusiveness of institutions is not new, and can be traced back to the work of scholars from the early twentieth century, like Harold Innis (1970), and to more recent scholars, including Engerman and Sokoloff (1997), or Acemoglu and Robinson (2012). Innis was a Canadian political economist who wrote extensively about how Canada's early exports (mainly fur) helped determine Canada's institutions (i.e. the Canadian relationships with Native Americans and with Europe). More recently, Engerman and Sokoloff (1997) and Acemoglu and Robinson (2012) have built an institutional theory of international differences in income based on the idea that colonial powers installed different institutions in their colonies. According to the theory, settlers installed extractive industries and institutions when they found unfavorable conditions, but created the cities that homed both, non-extractive activities and inclusive institutions, when they found favorable conditions that allowed them to migrate in mass.

From a modeling perspective, we can understand the co-evolution between productive structures, institutions, and human capital, by assuming a model of heterogeneous firms in which firms survive only when they are able to adopt or discover the institutions and human capital that work best in the industry that they participate in. This model assumes that institutions are to an significant extent created at work and depend on the type of industry. This assumption is extremely likely, because on the one hand, people learn to interact and collaborate with others in work settings, and on the other, there are clearly marked differences in the institutions (or culture) of different sectors. For instance, the liberal institutions that are common in Silicon Valley's tech sector might be ideal for industries that require workers to be creative problem solvers. These institutions, however, might be suboptimal in the context of a mining operation where following rules and respecting hierarchies can ensure the safety of workers and the coordination of the entire mining operation.

Of course, there are often differences in the ways different countries produce the same product. For instance, the same industry might be more labor-intensive in one country and more capital intensive in another country. However, there are also significant differences in the particular skills, knowledge, factor endowments and institutions that are needed to become globally competitive in a particular industry. For instance, the production of cocoa beans or coffee depends on the availability of natural resources, while the production of complex industries (like jet engines) depends on an extensive network of skilled workers. We do not know a priori whether differences among countries producing the same product (e.g. the production of cars in Spain and Japan), are larger or smaller than differences among the production of different products in the same country (the production of oranges and cars in Spain). Yet, the fact that we find a strong connection between productive structures and income inequality suggests—but does not prove—that differences among the processes required to produce the same product in different countries, may be smaller than the differences among the processes required to produce different products in the same countries.

Therefore we argue in this paper that countries exporting complex industries tend to be more inclusive and have lower levels of income inequality than countries that are exporting simpler products. Some countries may be able to achieve comparable high levels of average income based on natural resources, but those comparably high levels of income will rarely come with inclusive institutions when they are not the result of sophisticated industrial structures.

A country's productive structure can help explain variations in institutions and income inequality, however, only if there are significant differences in the productive structures of macro-economically similar countries. But how different are the productive structures of macro-economically similar countries? As an example consider Chile and Malaysia. In 2012, Chile's average income per capita and years of schooling ($21,044 at PPP in current 2012 US$ and 9.8 mean years of schooling) was comparable to Malaysia's income per capita and schooling ($22,314 and 9.5). Yet, their productive structures were practically orthogonal (see Figure 1), since Malaysia's exports mostly involved machinery and electronics, while Chile's exports mostly involved agriculture and mining. The Economic Complexity Index (ECI) captures these differences in productive structure. In 2012 Malaysia ranked 24th in the ECI ranking while Chile ranked only 72$^{nd}$ (for the ECI rankings and further information about export structures see atlas.media.mit.edu). Moreover, these differences in the ECI ranking also point more accurately to differences in these countries' level of income inequality. Chile's inequality as measured through the Gini coefficient (Gini$_{CHL}$=0.49) is significantly higher than that of Malaysia (Gini$_{MYS}$=0.39), illustrating the correlation between income inequality and productive structures (see also Figure 2). The remainder of the paper is dedicated to statistically testing this relationship for a large set of countries and years, as well as creating a product level index of income inequality that allows for the visualization of the co-evolution between the structural transformation and income inequality for all countries in our dataset.

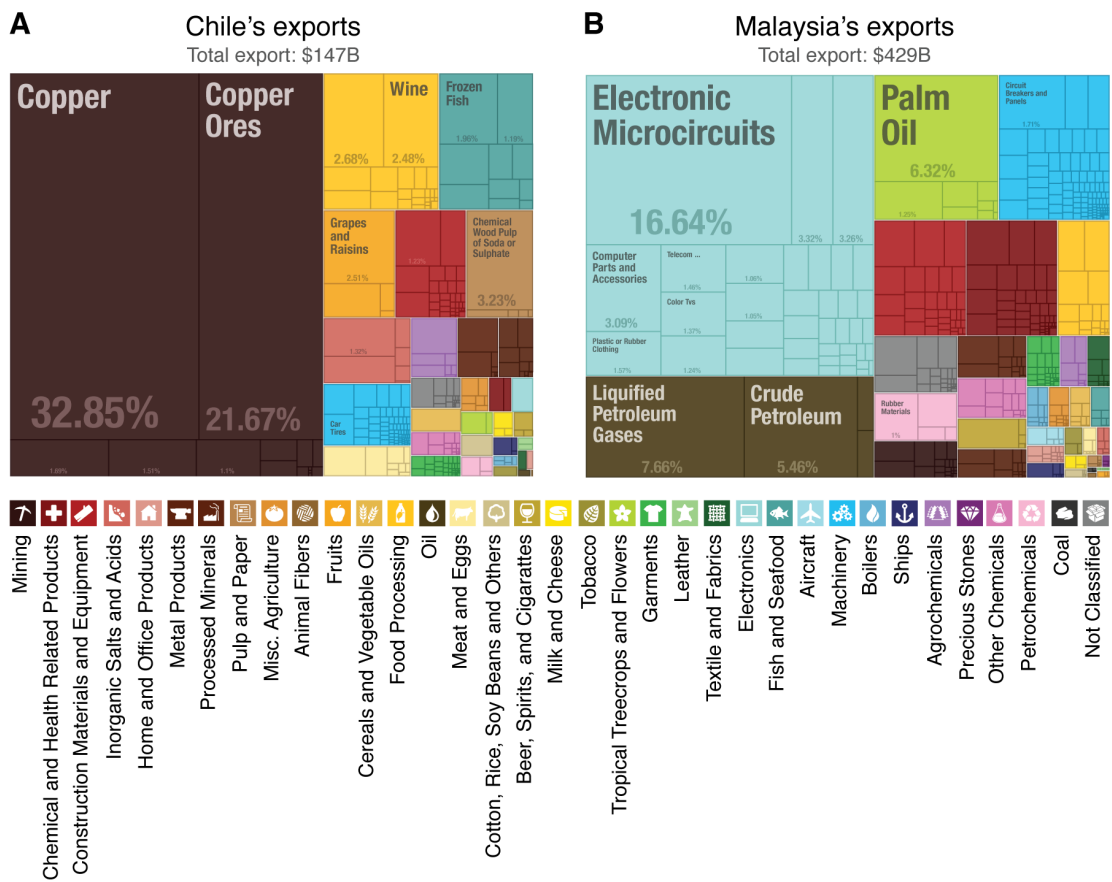

Figure 1. *Export structure of Chile (A) and Malaysia (B) in 2012.*
Source: atlas.media.mit.edu

3. DATA AND METHODS

We use data from the Economic Complexity Index (ECI), as well as international trade data, from MIT's Observatory of Economic Complexity (atlas.media.mit.edu) (Simoes & Hidalgo 2011). The trade data set combines exports data from 1962 to 2000, compiled by Feenstra et al. (2005), and data from the U.N. Comtrade for the period between 2001 and 2012.

The Economic Complexity Index (ECI) measures the sophistication of a country's productive structure by combining information on the diversity of a country (the number of products it exports), and the ubiquity of its products (the number of countries that export that product) (Hidalgo & Hausmann, 2009). The intuition behind ECI is that sophisticated economies are diverse and export products that, on average, have low ubiquity, because only a few diverse countries can make these sophisticated products. By the same token, less sophisticated economies are expected to produce a few ubiquitous products. ECI exploits this variation in the diversity of countries and the ubiquity of products to create a measure of a country's productive structure that incorporates information about the sophistication of products.

ECI is calculated from exports data connecting countries to the products in which they have Revealed Comparative Advantages (RCA) (Hidalgo & Hausmann, 2009). The Revealed Comparative Advantage (RCA) of a country $c$ in a product $p$ is:

$$RCA_{cp} = \frac{X_{cp} / \sum_{p'} X_{cp'}}{\sum_{c'} X_{c'p} / \sum_{c'p'} X_{c'p'}}$$

where $X_{cp}$ is the total export of country $c$ in product $p$. RCA is larger than 1 (indicating that a country has comparative advantage in a product), if a country's export of a product are larger than what would be expected from the size of the country's export economy and the product's global market.

RCA are used to define a discrete matrix $M_{cp}$ which is equal to 1 if country $c$ has RCA in product p and 0 otherwise.

$$M_{cp} = 1 \text{ if } RCA_{cp} \geq 1$$
$$M_{cp} = 0 \text{ if } RCA_{cp} < 1$$

The matrix $M_{cp}$ allows to define the diversity of a country and the ubiquity of a product, respectively, as the number of products that are exported by a country with comparative advantage, and the number of countries that export a product with comparative advantage.

$$Diversity = k_{c0} = \sum_p M_{cp}$$

$$Ubiquity = k_{p0} = \sum_c M_{cp}$$

Next, a matrix can be defined that connects countries exporting similar products, weighted by the inverse of the ubiquity of a product (to discount common products), and normalized by the diversity of a country:

$$\tilde{M}_{cc'} = \frac{1}{k_{c,0}} \sum_p \frac{M_{cp} M_{c'p}}{k_{p,0}}$$

Finally, the economic complexity index (ECI) is defined as

$$ECI_c = \frac{K_c - \langle K \rangle}{std(K)}$$

where $K_c$ is the eigenvector of $\tilde{M}_{cc'}$ associated with the second largest eigenvalue—the vector associated with the largest eigenvalue is a vector of ones (Hausmann et al., 2014; Caldarelli et al., 2012, Kemp-Benedict, 2014).

Table 1 shows the top five and bottom five economies based on the ranking provided by ECI for the year 2012.

Table 1. *Top and bottom 5 countries in the Economic Complexity Ranking 2012*

Top 5 countries by economic complexity

| ECI Rank | Country | GDP per capita (cons 2005 US$) | ECI |
|---|---|---|---|
| 1 | Japan | 36912 | 2.2288 |
| 2 | Switzerland | 58557 | 1.9734 |
| 3 | Germany | 39274 | 1.8396 |
| 4 | Sweden | 45260 | 1.7089 |
| 5 | South Korea | 23303 | 1.6999 |

Bottom 5 countries by economic complexity

| ECI Rank | ID | GDP per capita (cons 2005 US$) | ECI |
|---|---|---|---|
| 120 | Nigeria | 1034 | -1.6289 |
| 121 | Turkmenistan | 3270 | -1.7245 |
| 122 | Guinea | 303 | -1.7804 |
| 123 | Angola | 2426 | -2.2458 |
| 124 | Libya | 7078 | -3.1767 |

Source: atlas.media.mit.edu, World Bank's World Development Indicators

Our income inequality data comes from the Gini coefficient estimates by Galbraith et al., 2014) (GINI EHII dataset) and the GINI ALL dataset from Milanovic (2013). We mainly use the GINI EHII dataset because it is more comprehensive than other datasets (such as the GINI ALL dataset from Milanovic (2013) for the period between 1963 and 1989. The GINI EHII dataset contains inequality data for more than 60 countries starting in 1970, whereas the GINI ALL dataset contains data for less than 40 countries for periods between 1963 and 1989. Moreover the GINI ALL and is biased towards countries with high levels of GDP per capita and economic complexity (see figure A1 in the Appendix).

Additionally we use data on a country's *GDP per capita at Purchasing Power Parity (PPP) in constant 2005 US$*, *average years of schooling*, and *population*, from the World Bank's World Development Indicators. Data on institutional variables: *corruption control*, *political stability*, *government effectiveness*, *regulatory quality* and *voice and accountability*, come from the World Bank's Worldwide Governance Indicators (http://data.worldbank.org).

It must be noted that we consider only countries with a population larger than 1.5 million and total exports of over 1 billion dollars, thus removing small national economies that are comparable to medium-size cities. The resulting dataset includes 91% of the total world population and 84% of the total world trade between 1963 and 2008. Because of the sparseness of several variables, especially the Gini data, we use average values for the time periods 1963-1969, 1970-1979, 1980-1989, 1990-1999, and 2000-2008. We note that

Gini values change relatively slow, so these averages are close to the Gini values expected for each year within a decade. Summary statistics for all variables and each time period can be found in the Appendix (Table A1).

## 4 THE STATISTICAL CONNECTION BETWEEN ECONOMIC COMPLEXITY AND INCOME INEQUALITY

In this section we use bivariate and multivariate statistics to explore the connection between economic complexity and income inequality and to test its robustness to a number of controls.

### 4a. BIVARIATE ANALYSIS: COMPARING THE PREDICTIVE POWER OF DIFFERENT MEASURES OF PRODUCTIVE STRUCTURES

Figure 2 illustrates the bivariate relationship between Economic Complexity and Income Inequality in different decades, and compares the bivariate relationship between ECI-GINI with the bivariate relationship between ECI and GDP per capita (at Purchasing Power Parity, in constant 2005 US$) for 79 countries between 2000 and 2008. Both economic complexity and GDP per capita show a negative relationship with income inequality (Figure 2-A and 2-B). However, the negative relationship between economic complexity and income inequality ($R^2$=0.58, p-value=~$10^{-16}$) is stronger than the relationship between income inequality and GDP per capita ($R^2$=0.36, p-value= ~$10^{-10}$), and the difference in $R^2$ between these two bivariate regressions is statistically significant. To test for the significance of the difference in $R^2$ we used the Clarke-Test for non-nested models. Just as the F-test is the standard method to select among nested models, the Clarke-test is the basic statistical model used to test the significance of differences in $R^2$ for non-nested models (models where the independent variables are not perfect subsets of each other). In our case, the Clarke-Test prefers the ECI-GINI model over the GDP per capita-GINI model with a p-value=$4.3 \times 10^{-7}$ (also see Table A2 in the Appendix). Figures 2-C to 2-F, furthermore, show that the negative bivariate relationship between income inequality and economic complexity is stable across all considered decades. In all decades, we find a negative and significant relationship between economic complexity and income inequality.

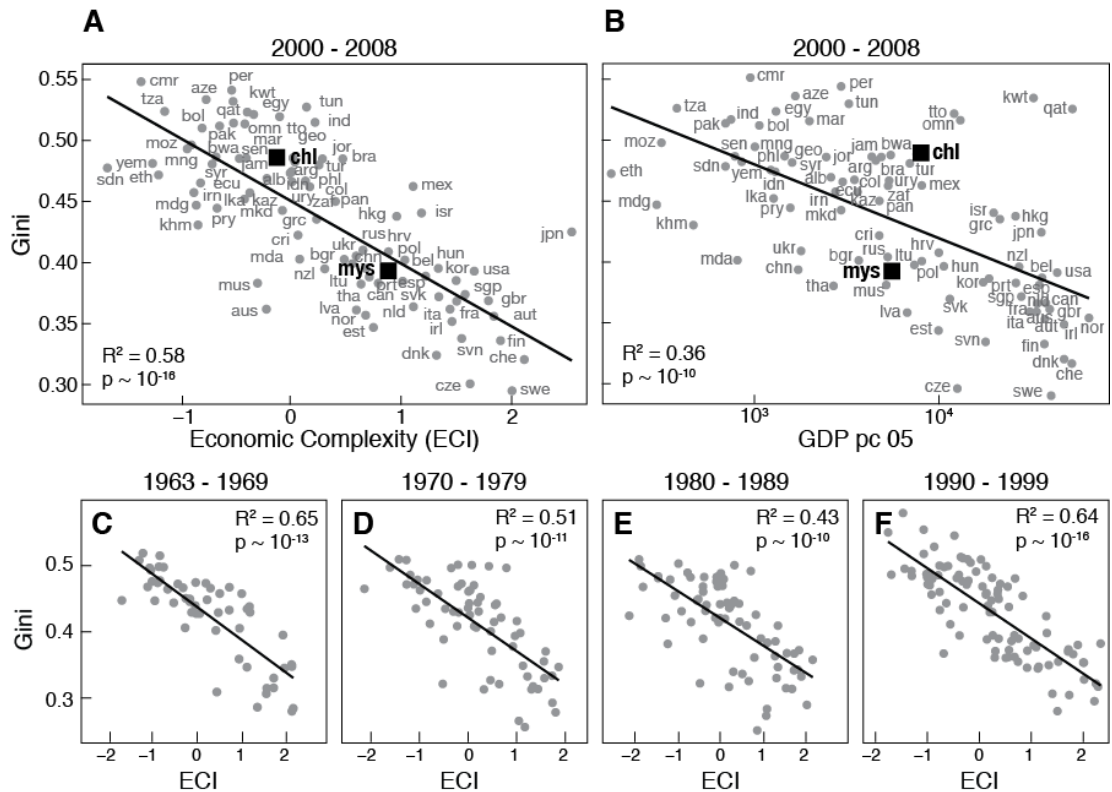

Figure 2. *Bivariate relationships between economic complexity, income, and income inequality.*

*Notes*: All figures show that $R^2$ and all p-values are less than $10^{-10}$. (**A**) ECI versus GINI EHII in 2000-2008. (**B**) Natural logarithm of GDP per capita (constant 2005 US$) versus GINI EHII. (**C**) ECI versus GINI EHII in 1963-1969, (**D**) 1970-1979, (**E**) 1980-1989 and (**F**) 1990-1999.

Next we compare the bivariate relationships between different measures of productive structures and income inequality in the period 2000-2008 (Figure 3). The matrix diagonal of Figure 3 illustrates the histograms of each variable, the upper triangle of the matrix shows the correlation coefficients between each pair of variables, and the lower triangle shows the corresponding scatterplots with a smoothed conditional mean line. ECI has strong and significant correlations with all other measures of productive structure, while ECI has a higher correlation with the income inequality measures GINI EHII and GINI ALL than GDP and all other measures of productive structures. Table A2 and Table A3 in the Appendix also show the result of the Clarke tests comparing the predictive power of ECI for income inequality with other measures of productive structures for time periods. In the case of GINI EHII data, ECI is significantly preferred as predictor variable in 15 out of 16 model comparisons, whereas only in one model comparison neither model is significantly preferred. In case of the GINI ALL data, ECI is significantly preferred in 13 out of 16 model comparisons, while in three cases neither model is preferred. There is no case which income per capita or measures of productive structures are significantly preferred as predictor variable for income inequality in comparison to ECI.

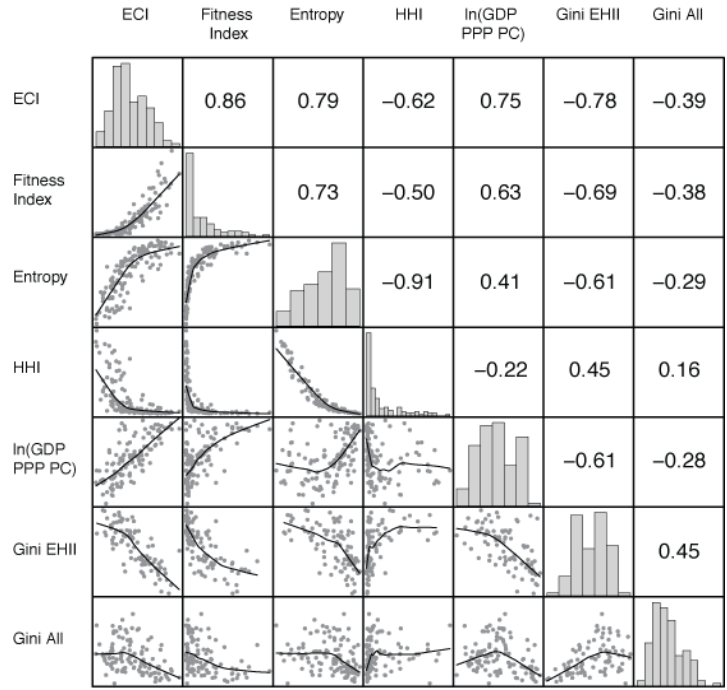

Figure 3. Correlations between different economic diversity measures and income inequality in 2000-2008. HHI refers to the Herfindahl-Hirschman Index and is a commonly used concentration measure; the Economic Complexity Index (ECI), the Fitness Index and Shannon Entropy are used to measure the diversity and sophistication of a country's exports.

Next, we proceed to cross-sectional and panel regressions to see if there is a significant correlation between economic complexity and inequality when controlling for other factors of inequality like human capital or institutions. Afterwards, we present a new measure that allows us to estimate the level of inequality related to different types of products. This measure, in combination with the network of related products (Hidalgo et al., 2007), shows how the productive structure constrains a country's income inequality and opportunities for inclusive economic development.

## 4b. MULTIVARIATE REGRESSION RESULTS

We start our analysis with a pooled regression for the period from 1996 to 2008, and then we explore the changes in Ginis, between 1960s and 2000s, using a panel regression for each decade that includes country fixed effects. Because of the sparseness of the Gini datasets and slow temporal changes in Ginis, we use average values for each panel. We use the periods 1996-2001 and 2002-2008 for cross-section regressions and 1963-1969, 1970-1979, 1980-1989, 1990-1999, and 2000-2008 for the fixed effects panel regression. Due to the sparseness of the institutional variables, we only include them in the cross-section regressions.

# POOLED REGRESSION

Table 2 shows a pooled cross-sectional regression for the periods of time between 1996-2001 and 2002-2008. Columns 1 to 6 illustrate a sequence of nested models that regress income inequality against *economic complexity*, *GDP per capita at Purchasing Power Parity (PPP)* and its square (a.k.a. Kuznets' Curve), *average years of schooling*, *population* and the institutional factors: *corruption control*, *government effectiveness*, *political stability*, *voice and accountability*, and *regulatory quality*.

In every model, the Economic Complexity Index (ECI) is a negative and significant predictor of income inequality. Education (as measured by average years of schooling) and log GDP squared also show a negative and significant correlation with inequality; log GDP a positive and significant correlation. Notably, when we control for economic complexity, then the rising part of the Kuznets curve is even more pronounced than without it. Also the role of education in terms of years of schooling becomes less important. Together, all variables explain 69.3% of the variance in income inequality among countries (Table 2, Column 1), but ECI is the most significant variable in the regression analysis, and it is also the variable that explains the largest percentage of the variance in income inequality after the effects of all other variables have been taken into account. The semi-partial correlation of ECI (the difference in $R^2$ between the full model and one in which only ECI was removed) is 8.1%, meaning that 8.1% of the variance in income inequality—which is not accounted for by institutional and macroeconomic variables—is explained by ECI (Table 2). Conversely the semi-partial correlations of all institutional variables is less than 0.1%, while that of income, population, and education, are all individually less than 2%. This means that these variables capture information about inequality that is already largely captured by ECI. Furthermore, ECI contains additional information about inequality that cannot be explained by these other variables alone.

Table 2. *Pooled OLS regression models*

| | *Dependent variable: Gini* | | | | | |
|---|---|---|---|---|---|---|
| | (I) | (II) | (III) | (IV) | (V) | (VI) |
| ECI | -0.040*** | | -0.037*** | -0.046*** | -0.033*** | -0.044*** |
| | (0.007) | | (0.007) | (0.007) | (0.006) | (0.006) |
| ln(GDP PPP pc) | 0.067** | 0.059* | | 0.060** | 0.056* | 0.075*** |
| | (0.028) | (0.032) | | (0.029) | (0.028) | (0.025) |
| ln(GDP PPP pc)$^2$ | -0.004** | -0.004* | | -0.003* | -0.003* | -0.004*** |
| | (0.002) | (0.002) | | (0.002) | (0.002) | (0.001) |
| Schooling | -0.005*** | -0.009*** | -0.004** | | -0.006*** | -0.005*** |
| | (0.002) | (0.002) | (0.002) | | (0.002) | (0.002) |
| Ln Population | 0.007** | 0.0001 | 0.005* | 0.008*** | | 0.009*** |
| | (0.003) | (0.003) | (0.003) | (0.003) | | (0.002) |
| Rule of law | -0.013 | -0.016 | -0.016 | -0.015 | -0.013 | |
| | (0.013) | (0.014) | (0.013) | (0.013) | (0.013) | |
| Corruption Control | 0.011 | 0.027* | 0.009 | 0.016 | 0.007 | |
| | (0.013) | (0.014) | (0.013) | (0.013) | (0.013) | |
| Government Effectiveness | 0.002 | -0.022 | 0.003 | 0.006 | 0.010 | |
| | (0.017) | (0.018) | (0.017) | (0.017) | (0.017) | |
| Political Stability | -0.010 | -0.017** | -0.009 | -0.009 | -0.017*** | |
| | (0.006) | (0.007) | (0.006) | (0.006) | (0.006) | |
| Regulatory Quality | -0.006 | -0.012 | -0.0002 | -0.010 | -0.012 | |
| | (0.012) | (0.014) | (0.012) | (0.012) | (0.012) | |
| Voice and accountability | 0.001 | 0.006 | 0.001 | -0.004 | 0.003 | |
| | (0.008) | (0.009) | (0.008) | (0.008) | (0.008) | |
| Constant | 0.083 | 0.286** | 0.391*** | 0.068 | 0.244** | 0.016 |
| | (0.130) | (0.141) | (0.050) | (0.134) | (0.114) | (0.121) |
| Observations | 142 | 142 | 142 | 142 | 142 | 142 |
| $R^2$ | 0.717 | 0.639 | 0.701 | 0.699 | 0.704 | 0.704 |
| Adjusted $R^2$ | 0.693 | 0.612 | 0.681 | 0.676 | 0.681 | 0.693 |
| Residual Std. Error | 0.035 (df = 130) | 0.039 (df = 131) | 0.035 (df = 132) | 0.035 (df = 131) | 0.035 (df = 131) | 0.035 (df = 136) |
| F-Statistic | 29.916*** (df = 11; 130) | 23.208*** (df = 10; 131) | 34.413*** (df = 9; 132) | 30.458*** (df = 10; 131) | 31.165*** (df = 10; 131) | 64.656*** (df = 5; 136) |

*Notes:* *p<0.1; **p<0.05; ***p<0.01
These pooled OLS regression models regress income inequality against economic complexity, a country's average level of income and its square, population, human capital and the institutional variables: rule of law, corruption control, government effectiveness, political stability, regulatory quality, and voice and accountability. Column I includes all variables. Columns II-VI exclude blocks of variables to explore the contribution of each group of variables to the full model. The sharpest drop in R2 (from 0.693 to 0.612) is observed when ECI is removed from the regression. The table pools data from two panels, one from 1996-2001 and another one from 2002-2008. The numbers in parenthesis are standard errors (SEM).

In the Tables A4-A8 of the Appendix we also test these results within each decade as well as using alternative Gini data sets (Milanovic, 2013; Galbraith et al., 2014) and alternative measures of economic diversity, concentration, and fitness (Hirschman, 1945; Herfindahl, 1950; Imbs & Wacziarg, 2003; Teece et al., 1994; Frenken, Oort & Verburg, 2007; Saviotti & Frenken, 2008; Boschma & Iammarino, 2009; Tacchella et al., 2012; Cristelli, Tacchella & Pietronero, 2015). We find that our results are robust to these changes in datasets, methods, and classifications.

## TEMPORAL CHANGES

Next, we explore whether changes in a country's level of economic complexity are associated with changes in income inequality by using a country-fixed-effect panel regression with decade panels from 1963 to 2008. Unlike cross-sectional results, which make use of variations in inequality between countries, fixed-effect panel regressions exploit temporal variations within a country. These variations are small for both income inequality and economic complexity, and thus we should not expect large effects. Yet, despite the low levels of temporal variation in the data, the fixed-effect panel regression still reveals a negative and significant association between a country's change in economic complexity and in its Gini coefficient (Table 3), meaning countries that experienced an increase in economic complexity tended to experience a decrease in income inequality. In fact, we find that an increase in one standard deviation in economic complexity is associated with a reduction in Gini of 0.03. This is equivalent to three extra years of schooling. This association between changes in economic complexity and income inequality is robust to the inclusion of measures of income and human capital. The institutional variables are not included since that data is only available for only one of the five panels (the most recent one).

Table 3. *Fixed-effects panel regression*

|  | *Dependent variable: GINI* | | | | | | |
|---|---|---|---|---|---|---|---|
|  | I | II | III | IV | V | VI | VII |
| ECI | -0.031*** | -0.033*** | -0.024*** | -0.026*** |  | -0.030*** | -0.029*** |
|  | -0.007 | -0.007 | -0.007 | -0.007 |  | -0.007 | -0.007 |
| ln(GDP PPP pc) |  | -0.038 | -0.042 | -0.017 | -0.032 |  | -0.053* |
|  |  | -0.028 | -0.027 | -0.029 | -0.03 |  | -0.03 |
| ln(GDP PPPpc)$^2$ |  | 0.003* | 0.002 | -0.00003 | 0.0005 |  | 0.004** |
|  |  | -0.002 | -0.002 | -0.002 | -0.002 |  | -0.002 |
| Schooling |  |  | 0.010*** | 0.014*** | 0.015*** | 0.010*** |  |
|  |  |  | -0.002 | -0.003 | -0.003 | -0.002 |  |
| Ln Population |  |  |  | -0.024** | -0.016 | -0.022** | 0.014* |
|  |  |  |  | -0.011 | -0.011 | -0.01 | -0.008 |
| Observations | 338 | 338 | 338 | 338 | 338 | 338 | 338 |
| R$^2$ | 0.077 | 0.123 | 0.198 | 0.213 | 0.165 | 0.196 | 0.134 |
| Adjusted R$^2$ | 0.055 | 0.087 | 0.139 | 0.149 | 0.116 | 0.138 | 0.094 |
| F-Statistic | 20.13*** (df = 1; 240) | 11.14*** (df = 3; 238) | 14.63*** (df = 4; 237) | 12.80*** (df = 5; 236) | 11.74*** (df = 4; 237) | 19.36*** (df = 3; 238) | 9.15*** (df = 4; 237) |
| Country Fixed Effects | Yes | Yes | Yes | Yes | Yes | Yes | Yes |

*Notes*: *p<0.1; **p<0.05; ***p<0.01
These seven fixed-effects panel regression models explore whether changes in a country's level of economic complexity are associated with changes in income inequality (column I), also controlling for the effects that other socioeconomic factors like income (column II), human capital (column III) and population (column IV) have on income inequality. Columns V-VII control the variance explained by the model when ECI, income, or schooling, are excluded from the analysis. The numbers in parenthesis are standard errors (SEM).

This shows that the mix of products that a country exports is a significant predictor of income inequality in both cross-sectional and panel regressions, even when controlling for other aggregated socioeconomic variables, like GDP, education or population. Naturally, future research could address variations within educational achievements or variables like within-countries differences in access to infrastructure in more detail (Acemoglu & Dell, 2010).

While in this section we controlled the significant general trends of the relationship between economic complexity and income inequality, the next section explores the inequality related to 775 particular product categories and visualizes the importance of the network structure of production—or product space— for the subsequent diversification of countries into more inclusive industries or less inclusive industries.

# 5. THE PRODUCT SPACE AND INCOME INEQUALITY

This section first estimates the expected inequality related to different types of products and then uses this information to visualize the country unique structural constraints on economic development and income inequality.

## DECOMPOSING INEQUALITY AT THE PRODUCT LEVEL

We decompose the relationship between economic complexity and income inequality into individual economic sectors by creating a product level estimator of the income inequality that is expected for the countries exporting a given product. We call this product level indicator the Product Gini Index (PGI). The PGI is closely related to previous measures on the sophistication of exports—i.e. the export sophistication measure of Lall, Weiss & Zhang
 (2006), the PRODY of Hausmann, Hwang & Rodrik (2006) and the Product Complexity Index of Hausmann and Hidalgo (2009)—showing that the type of products a country exports determines its level of economic development (Rodrik, 2006; Hausmann, Hwang & Rodrik, 2006; Hidalgo et al., 2007; Hidalgo and Hausmann, 2009). However, instead of relating products to income of the countries exporting these products, the PGI explores the association of products with different levels of income inequality.

Decomposing income inequality at the product level can be understood in the context of the co-evolution between productive structures, education, and institutions, as discussed in the introduction. To decompose income inequality at the product level we define the Product Gini Index (PGI) as the average level of income inequality of a product's exporters, weighted by the importance of each product in a country's export basket. Formally, we define the *PGI (Product Gini Index)* for a product $p$ as:

$$PGI_p = \frac{1}{N_p} \sum_c M_{cp} s_{cp} Gini_c \quad (1)$$

where $Gini_c$ is the Gini coefficient of country $c$, $M_{cp}$ is 1 if country $c$ exports product $p$ with revealed comparative advantage and 0 otherwise, $s_{cp}$ is the share of country $c$'s exports represented by product $p$. $N_p$ is a normalizing factor that ensures PGIs are the weighted average of the Ginis. $N_p$ and $s_{cp}$ are calculated as:

$$N_p = \sum_c M_{cp} s_{cp}$$

where,

$$s_{cp} = X_{cp} \Big/ \sum_{p'} X_{cp'}$$

where $X_{cp}$ is the total export of product $p$ by country $c$.

We estimate PGIs using an average of Ginis for each product, instead of using a regression with product dummies, because the number of products in our data is much larger than the number of countries (e.g. 775 vs. 92 in 1995-2008), and hence, a regression would be over-specified.

Figure 4A illustrates the construction of the PGI and Figure 4B shows the top 3, bottom 3 and median 3 products according to the ranking of PGI values between 1995 and 2008 (for all products see Table A12 in the Appendix). The products associated with the highest levels of income inequality (high PGI) mainly consist of commodities, such as Cocoa Beans, Inedible Flours of Meat and Fish, and Animal Hair. Low PGI products, on the other hand, include more sophisticated forms of machinery and manufacturing products, such as Paper Making Machine Parts, Textile Machinery, and Road Rollers.

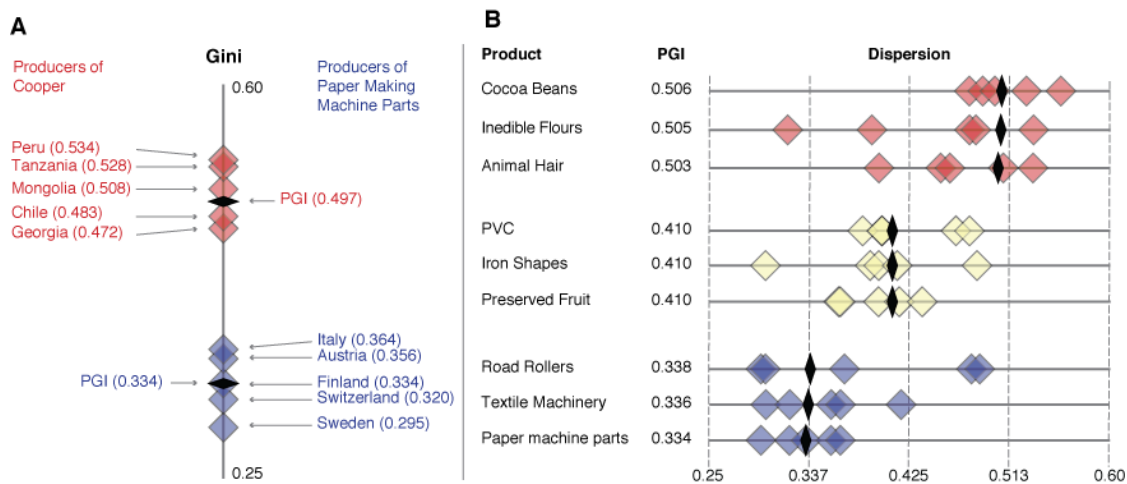

Figure 4. *The Product Gini Index (PGI)*.

*Notes*: (A). The product Gini index (PGI) is a weighted average of the Gini coefficients of the countries that export a product. The Gini coefficients of five copper exporters is shown in red. In blue, we show the Gini coefficients of exporters of paper making machine parts. (B). Top three, middle three, and bottom three products by PGI values. The PGI value is indicated with a black diamond. The Gini values of the five countries that contribute the most to each of these PGI are shown using diamonds. All values are measured using data from 1995-2008.

Further information and descriptive statistics about the Product Ginis can also be found in the Tables A9-A11 and Figure A2 of the Appendix. It must be noted that products with a high level of economic complexity—measured by the product complexity index (Hidalgo & Hausmann, 2009; Felipe et al., 2012; Hausmann et al. 2014)—tend to also have lower PGI values (Figure A2 of the Appendix). This means that complex products—such machine parts or electronic equipment for industrial chains or I-Phones, robots or 3D printing devices—tend to be produced in more equalitarian countries than simpler and resource-exploiting products like cocoa beans or copper. It is common sense that complex products require a larger network of skilled workers, related industries and inclusive institutions making the economic competitiveness of these products possible, than simpler industrial products and resource exploiting activities whose competitiveness is mainly based on resource richness, low labor costs, routinized activities and economies of scale. A related observation, which is important but beyond the scope of this paper is that these simpler products also tend to be located at the beginning or the end of global production chains—they are either extractive or assembly activities.

Naturally, the association between product complexity and income inequality also implies the need to understand the systemic distribution of inequality across the network of related products—or global product space (Hidalgo et al., 2007), and how productive transformations of the productive matrix are associated with changes in income inequality.

THE PRODUCT SPACE AND THE EVOLUTION OF INCOME INEQUALITY

In this section, we use PGIs in combination with the product space—the network connecting products that are likely to be co-exported—to show how changes in a country's productive structure are connected to changes in a country's level of income inequality.

Figure 5A assigns colors for each product, using PGIs between 1995 and 2008. Products associated with low levels of inequality (low PGIs) are located in the center of the product space, where the more sophisticated products are located. On the other hand, high PGI products tend to be located in the periphery of the product space, where less sophisticated products are located (6).

We can also use the product space to study the constraints on industrial diversification and the evolution of income inequality implied by a country's productive structure. The product space captures the notion that countries, cities, and regions, are significantly more likely to diversify towards products that are similar (i.e. connected in the product space) to the products that they currently export (Frenken, Oort & Verburg, 2007; Saviotti & Frenken, 2008; Boschma & Iammarino, 2009; Hidalgo et al., 2007; Hidalgo & Hausmann, 2009; Neffke, Henning & Boschma, 2011; Hausmann et al., 2014)

Figure 5B-G compares the evolution of the productive structure of Malaysia (5B-C), Norway (5D-E), and Chile (5F-G). Malaysia's economy evolved from high PGI products in 1963-1969—e.g. natural rubber and sawlogs—to low PGI products in 2000-2008—e.g. electronic microcircuits and computer parts. Norway, on the other hand, moved in the opposite direction, increasing its dependency on a high PGI product—crude petroleum—and saw an increase in income inequality. Finally, Chile developed in a more constrained way, diversifying into products with a relatively high PGI—frozen fish, fresh fish, and wine. More generally, these examples illustrate how the productive structure of a country constrains the evolution of its income inequality.

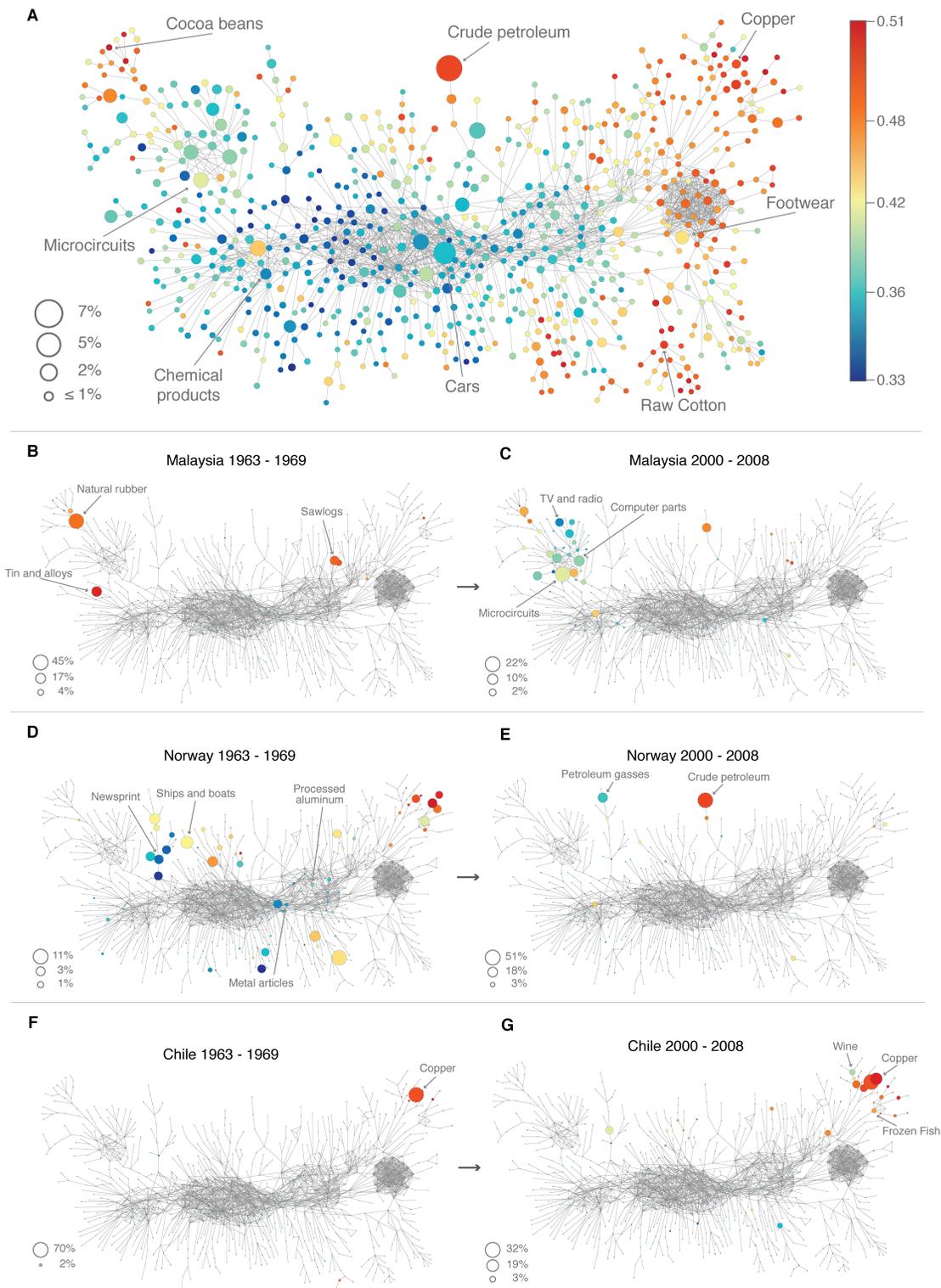

**Figure 5.** The product space and income inequality. (A) In this visualization of the product space nodes are colored according to a product's PGI as measured between 1995-2008. Node sizes are proportional to world trade between 2000 and 2008. The networks are based on a proximity matrix representing 775 SITC-4 product classes exported between 1963-2008. The link strength (proximity) is based on the conditional probability that the products are co-exported. (B) Malaysia's export portfolio between 1963-1969. In this figure and the subsequent ones node sizes indicate the share of a product in a country's export basket. Only products with RCA greater than 1 are presented. (C) Malaysia's export portfolio between 2000-2008. (D) Norway's exports between 1963-1969 and (E) between 2000-2008. (F) Chile's exports between 1963-1969 and (G) between 2000-2008.

## FACILITATING THE INTERACTIVE STUDY OF COUNTRIES' PATHS OF INCLUSIVE GROWTH

Data on all countries can be explored in an interactive web-tool we designed for this paper. The web-tool is available at MIT's observatory of economic complexity (atlas.media.mit.edu). The purpose of the interactive, visualization tool is to enable discussions among academicians, policy-makers and practitioners about inclusive growth opportunities that takes into account each country's unique productive structure, particular opportunities for inclusive growth, and historical-structural-developments. The interactive online tool allows the user to explore a map of 774 products between the years 1963 and 2008.

## DISCUSSION / CONCLUDING REMARKS

Our results illustrate that the ability of an economy to both generate and distribute income is strongly correlated with the mix of products a country is able to produce and export.

Taking economic complexity and productive space dynamics into account allows us to reveal structural linkages between economic development and income inequality which aggregated variables like the average years of schooling or income per capita alone are incapable of revealing. Our empirical results document a strong and robust correlation between the economic complexity index and income inequality. Using multivariate regression, we confirmed that this relationship is robust to controlling for measures of income, education, and institutions, and that this relationship has remained strong over the last fifty years. Moreover, we showed that increases in economic complexity tend to be accompanied by decreases in income inequality.

Our findings do not mean that productive structures solely determine a country's level of income inequality. On the contrary, a more likely explanation of the association between a country's productive structure and income inequality is that, as we argued in the introduction, productive structures represent a high-resolution expression of a number of factors, from institutions to education, that co-evolve with a country's mix of exported products and with the inclusiveness of its economy. Still, because of this co-evolution, our findings emphasize the economic importance of productive structures, since we have shown that these are not only associated with income and economic growth (Hidalgo et al., 2007; Hidalgo & Hausmann, 2009; Hausmann et al., 2014), but also with how income is distributed.

Moreover, we advance methods that enable a more fine-grained perspective on the relationship between productive structures and income inequality. The method is based on introducing the Product Gini Index or PGI, which estimates the expected level of inequality for the countries exporting a given product. Overlaying PGI values on the network of related products allows us to create maps that can be used to anticipate how changes in a country's productive structure could affect its level of income inequality. These maps provide means for researchers and policy-makers to explore and compare the co-evolution of productive structures, institutions and income inequality for hundreds of economies.

This paper and the related online tool show that a country's productive structure conditions its path of economic development and its abilities to generate and distribute income. This also implies that social and industrial policies may need to complement each other to achieve sustained inequality reduction and economic development (Stiglitz, 1996; Ranis, Steward & Ramirez, 2000; Amsden, 2010; Hartmann, 2014). While it is important for economic development and inequality reduction to improve school education and health services, it is also important to create advanced products and jobs that demand specialized education and inclusive institutions.

Of course, much more theoretical work needs to be done on the complex relationships between economic complexity, institutions and income inequality. For example, there is a need for more research on the importance of sector-wide institutions versus national institutions in different types of products. Moreover, the role of unions and different types of knowledge—e.g. tacit or codified knowledge— in the relationship between a country's productive matrix and its level of income inequality needs to be more thoroughly explored. Finally, the association between industrialization and gender equality requires further research.

More in-depth case studies are also required, for example in countries like Mexico and Australia. Mexico shows both a high level of income inequality and a high ECI. This might be due to the impact of US production facilities and a slightly inflated ECI value, a fact that is partly expressed in the lack of diversity of Mexico's export destinations. Conversely, Australia has partly succeeded in establishing inclusive institutions with relatively low levels of income inequality, despite having an export structure focused on natural resource extraction. Yet, there may be other activities that are not captured in trade data—such as the export of mining services and operations—that may help partly explain the anomaly. In other words, Australia might be a case where the measure of complexity based on exports underestimates the country's real complexity.

We must also note that our results are limited to the time period between 1963 to 2008. Yet, significant root causes of the present institutions, world production system, and distribution of income can be traced back to the Commercial Revolution and colonization in the 15$^{th}$ -18$^{th}$ century, as well as the Industrial Revolution that has started in the 18$^{th}$ century and has subsequently spread across the world. Certainly not all segments of the population have always benefitted from an increase in economic complexity, since many individuals suffered—or continue to suffer—from unemployment, slavery, or exploitative working conditions. Nonetheless, the social institutions necessary to allow countries to use new technologies, sustain creative destruction processes, and achieve high levels of economic complexity over the middle to long-run, have tended to include larger segments of the population back again into the economic development processes, at least in the industrially most advanced economies (Perez, 2003; Acemoglu & Robinson, 2012).

Despite the limitations, we showed that applying methods from economic complexity and network science allows us to capture significant information about countries' path of economic development that goes beyond aggregated input or output factors like GDP or years of schooling. The development pioneers highlighted the role of productive structures and structural transformations in economic development. Here we have introduced methods that are capable of exploring how heterogeneous productive structures of countries condition their ability to generate and distribute income in more detail. Moreover, we can use these methods to further explore the association between economic complexity and the evolution of institutions.

In sum, our analysis strongly suggests that countries exporting more complex products tend to have significantly lower levels of income inequality than countries exporting simple products.

# REFERENCES

1. Abdon, A., & Felipe, J. (2011). *The Product Space: What Does It Say About the Opportunities for Growth and Structural Transformation of Sub-Saharan Africa?* (Economics Working Paper Archive No. wp_670). Levy Economics Institute. Retrieved from https://ideas.repec.org/p/lev/wrkpap/wp_670.html
2. Acemoglu, D., Aghion, P., & Violante, G. L. (2001). *Deunionization, Technical Change and Inequality* (SSRN Scholarly Paper No. ID 267264). Rochester, NY: Social Science Research Network. Retrieved from http://papers.ssrn.com/abstract=267264
3. Acemoglu, D., & Dell, M. (2010). Productivity Differences between and within Countries. *American Economic Journal: Macroeconomics*, *2*(1): 169–88.
4. Acemoglu, D., & Robinson, J. (2012). *Why Nations Fail: The Origins of Power, Prosperity, and Poverty*. New York: Crown.
5. Amsden, A.H. (2010). Say's Law, Poverty Persistence, and Employment Neglect. Journal of Human Development and Capabilities 11(1), 57–66.
6. Autor, D. H. (2014). Skills, education, and the rise of earnings inequality among the "other 99 percent." *Science*, *344*(6186), 843–851. http://doi.org/10.1126/science.1251868
7. Beinhocker, E. D. (2006). *The Origin of Wealth: Evolution, Complexity, and the Radical Remaking of Economics*. Harvard Business Press.
8. Boschma, R., & Iammarino, S. (2009). Related variety, trade linkages, and regional growth in Italy. *Economic Geography*, *85*(3), 289–311. http://doi.org/10.1111/j.1944-8287.2009.01034.x
9. Bourguignon, F. (2004). *The Poverty-growth-inequality triangle* (Indian Council for Research on International Economic Relations, New Delhi Working Paper No. 125). Indian Council for Research on International Economic Relations, New Delhi, India. Retrieved from https://ideas.repec.org/p/ind/icrier/125.html
10. Brynjolfsson, E., & McAfee, A. (2012). *Race Against the Machine: How the Digital Revolution is Accelerating Innovation, Driving Productivity, and Irreversibly Transforming Employment and the Economy*. Lexington, MA: Digital Frontier Press.
11. Bustos, S., Gomez, C., Hausmann, R., & Hidalgo, C. A. (2012). The Dynamics of nestedness predicts the evolution of industrial ecosystems. *PLoS One*, *7*(11), e49393. http://doi.org/10.1371/journal.pone.0049393
12. Caldarelli, G., Cristelli, M., Gabrielli, A., Pietronero, L., Scala, A., & Tacchella, A. (2012). A Network Analysis of Countries' Export Flows: Firm Grounds for the Building Blocks of the Economy. *PLOS ONE*, *7*(10), e47278. http://doi.org/10.1371/journal.pone.0047278
13. Collier, P. (2007). *The Bottom Billion: Why the Poorest Countries are Failing and What Can Be Done About It*. New York: Oxford Univ. Press.
14. Cristelli, M., Gabrielli, A., Tacchella, A., Caldarelli, G., & Pietronero, L. (2013). Measuring the Intangibles: A Metrics for the Economic Complexity of Countries and Products. *PLOS ONE*, *8*(8), e70726. http://doi.org/10.1371/journal.pone.0070726
15. Cristelli, M., Tacchella, A., & Pietronero, L. (2015). The heterogeneous dynamics of economic complexity. *PLoS ONE*, *10*(2), e0117174. http://doi.org/10.1371/journal.pone.0117174

16. Davis, G. F. (2009). *Managed by the Markets: How Finance Re-Shaped America*. Oxford: Oxford Univ. Press.
17. Deininger, K., & Squire, L. (1998). New ways of looking at old issues: inequality and growth. *Journal of Development Economics*, *57*(2), 259–287. http://doi.org/10.1016/S0304-3878(98)00099-6
18. Engerman, S. L., & Sokoloff, K. L. (1997). Factor endowments, institutions, and differential paths of growth among new world economies. In S. H. Haber (Ed.), *How Latin America Fell Behind: Essays on the Economic Histories of Brazil and Mexico, 1800-1914* (pp. 260–304). California: Stanford Univ. Press.
19. Fajnzylber, F. (1990). Industrialization in Latin America: from the 'Black Box' to the 'Empty Box': a Comparison of Contemporary Industrialization Patterns. *Cuadernos de la CEPAL, No. 60, 1990-08*, Retrieved from http://hdl.handle.net/11362/27811
20. Feenstra, R. C., Lipsey, R. E., Deng, H., Ma, A. C., & Mo, H. (2005). *World Trade Flows: 1962-2000* (Working Paper No. 11040). National Bureau of Economic Research. Retrieved from http://www.nber.org/papers/w11040
21. Felipe, J. (2009). *Inclusive Growth, Full Employment and Structural Change: Implications and Policies for Developing Asia*. London: Anthem Press.
22. Felipe, J., Kumar, U., Abdon, A., & Bacate, M. (2012). Product complexity and economic development. *Structural Change and Economic Dynamics*, *23*(1), 36–68. http://doi.org/10.1016/j.strueco.2011.08.003
23. Frenken, K., Oort, F. V., & Verburg, T. (2007). Related variety, unrelated variety and regional economic growth. *Regional Studies*, *41*(5), 685–697. http://doi.org/10.1080/00343400601120296
24. Frey, C. B., & Osborne, M. A. (2013). *The Future of Employment: How susceptible are jobs to computerisation?* Oxford Martin School. Retrieved from http://www.oxfordmartin.ox.ac.uk/publications/view/1314
25. Fukuyama, F. (1996). *Trust: Human Nature and the Reconstitution of Social Order*. New York: Free Press.
26. Galbraith, J. K., Halbach, Béatrice, Malinowska, A., Shams, Amin, & Zhang, W. (2014). *UTIP Global Inequality Data Sets 1963-2008: Updates, Revisions and Quality Checks* (Working Paper No. 68). University of Texas at Austin. Retrieved from http://utip.gov.utexas.edu/papers/UTIP_68.pdf
27. Gustafsson, B., & Johansson, M. (1999). In Search of Smoking Guns: What Makes Income Inequality Vary over Time in Different Countries? *American Sociological Review*, *64*(4), 585–605. http://doi.org/10.2307/2657258
28. Hartmann, D. (2014). *Economic Complexity and Human Development: How Economic Diversification and Social Networks Affect Human Agency and Welfare*. New York: Routledge.
29. Hausmann, R., Hidalgo, C. A., Bustos, S., Coscia, M., Simoes, A., & Yildirim, M. A. (2014). *The Atlas of Economic Complexity: Mapping Paths to Prosperity*. MIT Press.
30. Hausmann, R., Hwang, J., & Rodrik, D. (2006). What you export matters. *Journal of Economic Growth*, *12*(1), 1–25. http://doi.org/10.1007/s10887-006-9009-4
31. Hausmann, R., & Rodrik, D. (2003). Economic development as self-discovery. *Journal of Development Economics*, *72*(2), 603–633. http://doi.org/10.1016/S0304-3878(03)00124-X
32. Herfindahl, O. C. (1950). *Concentration in the steel industry*.

33. Hidalgo, C. (2015). *Why Information Grows: The Evolution of Order, from Atoms to Economies*. New York: Penguin Press.
34. Hidalgo, C. A., & Hausmann, R. (2009). The building blocks of economic complexity. *Proceedings of the National Academy of Sciences*, *106*(26), 10570–10575. http://doi.org/10.1073/pnas.0900943106
35. Hidalgo, C. A., Klinger, B., Barabási, A.-L., & Hausmann, R. (2007). The product space conditions the development of nations. *Science*, *317*(5837), 482–487. http://doi.org/10.1126/science.1144581
36. Hirschman, A. O. (1945). *National power and the structure of foreign trade*. Berkeley and Los Angeles: University of California Press.
37. Hirschman, A. O. (1958). *The Strategy of Economic Development* (Vol. 10). New Haven: Yale Univ. Press.
38. Imbs, J., & Wacziarg, R. (2003). Stages of diversification. *The American Economic Review*, *93*(1), 63–86.
39. Innis, H. A. (1970). *The Fur Trade in Canada: An Introduction to Canadian Economic History*. Univ. of Toronto Press.
40. Kemp-Benedict, E. (2014). *An interpretation and critique of the Method of Reflections* (No. 60705). Retrieved from https://www.sei-international.org/publications?pid=2636
41. Kuznets, S. (1934). National Income, 1929-1932. In *National Income, 1929-1932* (pp. 1–12). NBER.
42. Kuznets, S. (1955). Economic growth and income inequality. *The American Economic Review*, *45*(1), 1–28.
43. Kuznets, S. (1973). Modern economic growth: findings and reflections. *The American Economic Review*, *63*(3), 247–258.
44. Lall, S., Weiss, J., & Zhang, J. (2006). The "sophistication" of exports: A new trade measure. *World Development*, *34*(2), 222–237. http://doi.org/10.1016/j.worlddev.2005.09.002
45. Leontief, W. W. (1951). Input-output economics. *Scientific American*, *185*(4), 15–21.
46. Milanovic, B. (2012). *Global Income Inequality by the Numbers: In History and Now - An Overview-*. The World Bank. Retrieved from http://elibrary.worldbank.org/doi/abs/10.1596/1813-9450-6259
47. Milanovic, B. L. (2013). All the Ginis Dataset. Retrieved June 28, 2015, from http://data.worldbank.org/data-catalog/all-the-ginis
48. MIT´s Observatory of Economic Complexity: Country Ranking. (n.d.). Retrieved June 28, 2015, from https://atlas.media.mit.edu/rankings
49. Neffke, F., Henning, M., & Boschma, R. (2011). How do regions diversify over time? Industry relatedness and the development of new growth paths in regions. *Economic Geography*, *87*(3), 237–265. http://doi.org/10.1111/j.1944-8287.2011.01121.x
50. Palma, J. G. (2011). Homogeneous middles vs. heterogeneous tails, and the end of the "Inverted-U": It's all about the share of the rich. *Development and Change*, *42*(1), 87–153. http://doi.org/10.1111/j.1467-7660.2011.01694.x
51. Perez, C. (2003). *Technological Revolutions and Financial Capital: The Dynamics of Bubbles and Golden Ages*. Cheltenham, UK; Northampton, MA, USA: Edward Elgar Pub.
52. Piketty, T. (2014). *Capital in the Twenty-First Century*. (A. Goldhammer, Trans.). Cambridge, MA: Belknap Press.

53. Ranis, G., Stewart, F. & Ramirez, A. (2000). Economic Growth and Human Development'. *World Development, 28* (2): 197–219.
54. Ravallion, M. (2004). *Pro-Poor Growth: A Primer* (Policy Research Working Paper No. ID 610283). World Bank. Retrieved from http://dx.doi.org/10.1596/1813-9450-3242
55. Rodrik, D. (2006). What's so special about China's exports? *China & World Economy*, *14*(5), 1–19. http://doi.org/10.1111/j.1749-124X.2006.00038.x
56. Rosenstein-Rodan, P. N. (1943). Problems of industrialisation of eastern and south-eastern Europe. *The Economic Journal*, *53*(210/211), 202–211. http://doi.org/10.2307/2226317
57. Sachs, J. (2005). *The End of Poverty: How We Can Make it Happen in Our Lifetime*. New York: Penguin Books Limited.
58. Saviotti, P. P., & Frenken, K. (2008). Export variety and the economic performance of countries. *Journal of Evolutionary Economics*, *18*(2), 201–218. http://doi.org/10.1007/s00191-007-0081-5
59. Shannon, C. E. (1948). A mathematical theory of communication. *The Bell System Technical Journal*, *27*, 379-423-656.
60. Singer, H. W. (1950). The distribution of gains between investing and borrowing countries. *The American Economic Review*, *40*(2), 473–485.
61. Stiglitz, J. E. (1996). Some lessons from the East Asian miracle. *The World Bank Research Observer*, *11*(2), 151–177. http://doi.org/10.1093/wbro/11.2.151
62. Stiglitz, J. E. (2013). *The Price of Inequality: How Today's Divided Society Endangers Our Future* (1 edition). New York: W. W. Norton.
63. Stiglitz, J., Sen, A. K., & Fitoussi, J.-P. (2009, December). The measurement of economic performance and social progress revisited: reflections and overview. Retrieved from https://hal-sciencespo.archives-ouvertes.fr/hal-01069384/document
64. Teece, D., Rumelt, R., Dosi, G., & Winter, S. (1994). Understanding corporate coherence: Theory and evidence. *Journal of Economic Behavior & Organization*, *23*(1), 1–30

APPENDIX

DESCRIPTIVE STATISTICS

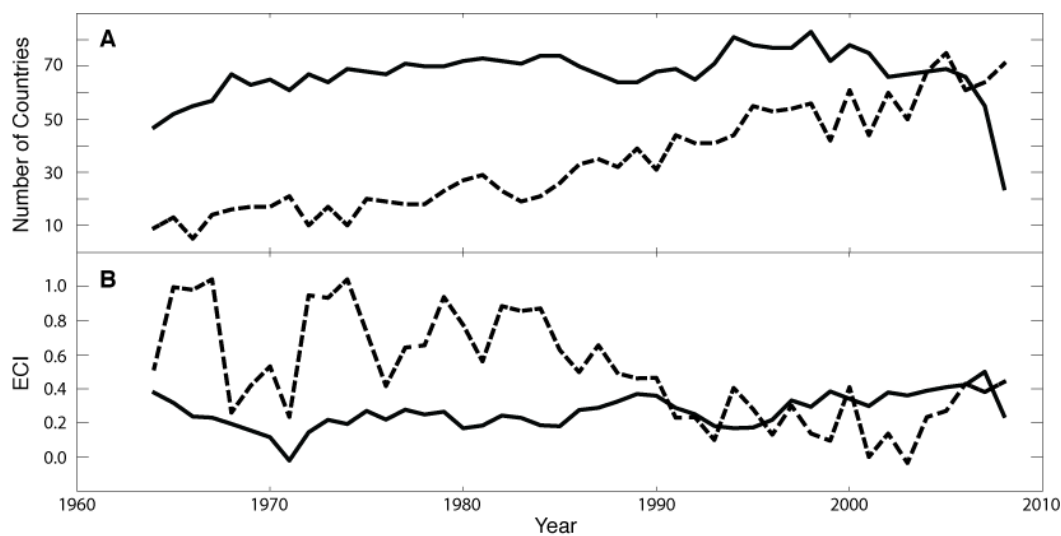

Figure A1. *Comparison between GINI EHII dataset (solid line) and GINI ALL dataset (dashed line).*

*Notes*: (A) Shows the total number of countries in each dataset, by year. (B) Shows average ECI of countries in each dataset, by year.

Table A1. *Descriptive summary statistics*

| Variable | Year | Obs. | Mean | SD | Min | Max |
| --- | --- | --- | --- | --- | --- | --- |
| ECI | 1963-1969 | 102 | 0.01 | 1.11 | -2.67 | 2.16 |
| ECI | 1970-1979 | 108 | -0.03 | 1.03 | -2.51 | 1.90 |
| ECI | 1980-1989 | 103 | 0.00 | 1.07 | -2.17 | 2.18 |
| ECI | 1990-1999 | 125 | 0.02 | 1.02 | -1.79 | 2.31 |
| ECI | 2000-2008 | 128 | 0.00 | 0.98 | -1.75 | 2.54 |
| ECI | All decades | 566 | 0.00 | 1.03 | -2.67 | 2.54 |
| GINI EHII | 1963-1969 | 85 | 42.67 | 7.35 | 22.32 | 54.59 |
| GINI EHII | 1970-1979 | 108 | 41.65 | 7.66 | 21.22 | 52.49 |
| GINI EHII | 1980-1989 | 121 | 41.48 | 7.59 | 20.93 | 53.09 |
| GINI EHII | 1990-1999 | 125 | 43.49 | 6.94 | 28.10 | 58.25 |
| GINI EHII | 2000-2008 | 103 | 43.69 | 6.40 | 29.62 | 55.02 |
| GINI EHII | All decades | 542 | 42.59 | 7.24 | 20.93 | 58.25 |
| GINI All | 1963-1969 | 44 | 41.46 | 10.06 | 20.88 | 62.00 |
| GINI All | 1970-1979 | 62 | 40.30 | 9.05 | 21.80 | 61.25 |
| GINI All | 1980-1989 | 99 | 36.31 | 11.24 | 18.60 | 62.90 |
| GINI All | 1990-1999 | 136 | 40.52 | 10.23 | 20.49 | 74.30 |
| GINI All | 2000-2008 | 148 | 40.05 | 9.55 | 23.98 | 72.95 |
| GINI All | All decades | 489 | 39.58 | 10.19 | 18.60 | 74.30 |
| GDP PPP05 pc | 1963-1969 | 71 | 4706.9 | 5984.5 | 113.0 | 21871.4 |
| GDP PPP05 pc | 1970-1979 | 76 | 6423.5 | 7881.9 | 168.8 | 29260.0 |
| GDP PPP05 pc | 1980-1989 | 92 | 8236.6 | 11578.5 | 150.1 | 59411.1 |
| GDP PPP05 pc | 1990-1999 | 115 | 8603.9 | 12061.4 | 128.2 | 51800.0 |
| GDP PPP05 pc | 2000-2008 | 123 | 11030.6 | 14669.9 | 157.6 | 64311.1 |
| GDP PPP05 pc | All decades | 477 | 8231.4 | 11624.0 | 113.0 | 64311.1 |
| ln(GDP PPP05 pc) | 1963-1969 | 71 | 7.64 | 1.34 | 4.73 | 9.99 |
| ln(GDP PPP05 pc) | 1970-1979 | 76 | 7.93 | 1.39 | 5.13 | 10.28 |
| ln(GDP PPP05 pc) | 1980-1989 | 92 | 7.97 | 1.56 | 5.01 | 10.99 |
| ln(GDP PPP05 pc) | 1990-1999 | 115 | 8.04 | 1.52 | 4.85 | 10.86 |

| Variable | Year | Obs. | Mean | SD | Min | Max |
| --- | --- | --- | --- | --- | --- | --- |
| ln(GDP PPP05 pc) | 2000-2008 | 123 | 8.33 | 1.51 | 5.06 | 11.07 |
| ln(GDP PPP05 pc) | All decades | 477 | 8.02 | 1.49 | 4.73 | 11.07 |
| ln(GDP PPP05 pc)2 | 1963-1969 | 71 | 60.08 | 20.83 | 22.35 | 99.86 |
| ln(GDP PPP05 pc)2 | 1970-1979 | 76 | 64.76 | 22.20 | 26.30 | 105.76 |
| ln(GDP PPP05 pc)2 | 1980-1989 | 92 | 65.94 | 25.15 | 25.11 | 120.83 |
| ln(GDP PPP05 pc)2 | 1990-1999 | 115 | 66.86 | 24.88 | 23.56 | 117.83 |
| ln(GDP PPP05 pc)2 | 2000-2008 | 123 | 71.68 | 25.35 | 25.60 | 122.58 |
| ln(GDP PPP05 pc)2 | All decades | 477 | 66.58 | 24.27 | 22.35 | 122.58 |
| Years of schooling | 1963-1969 | 81 | 3.63 | 2.56 | 0.44 | 10.17 |
| Years of schooling | 1970-1979 | 85 | 4.16 | 2.72 | 0.64 | 11.16 |
| Years of schooling | 1980-1989 | 86 | 4.85 | 2.82 | 0.85 | 11.80 |
| Years of schooling | 1990-1999 | 102 | 6.35 | 2.86 | 0.66 | 12.26 |
| Years of schooling | 2000-2008 | 105 | 7.32 | 2.84 | 0.88 | 12.92 |
| Years of schooling | All decades | 459 | 5.41 | 3.09 | 0.44 | 12.92 |
| Population (million) | 1963-1969 | 93 | 67.0 | 361.9 | 1.1 | 3401.4 |
| Population (million) | 1970-1979 | 96 | 77.9 | 422.6 | 1.0 | 4027.0 |
| Population (million) | 1980-1989 | 101 | 87.2 | 468.1 | 1.0 | 4560.0 |
| Population (million) | 1990-1999 | 119 | 44.8 | 142.0 | 1.1 | 1196.0 |
| Population (million) | 2000-2008 | 125 | 48.2 | 155.1 | 1.0 | 1294.4 |
| Corruption control | 1990-1999 | 110 | 0.08 | 1.06 | -1.28 | 2.40 |
| Corruption control | 2000-2008 | 116 | 0.08 | 1.03 | -1.48 | 2.48 |
| Corruption control | All decades | 226 | 0.08 | 1.04 | -1.48 | 2.48 |
| Government effectiveness | 1990-1999 | 110 | 0.13 | 0.97 | -1.25 | 2.12 |
| Government effectiveness | 2000-2008 | 116 | 0.15 | 0.97 | -1.51 | 2.18 |
| Government effectiveness | All decades | 226 | 0.14 | 0.97 | -1.51 | 2.18 |
| Political stability | 1990-1999 | 110 | -0.05 | 0.90 | -2.38 | 1.51 |
| Political stability | 2000-2008 | 116 | -0.06 | 0.89 | -2.08 | 1.58 |
| Political stability | All decades | 226 | -0.05 | 0.89 | -2.38 | 1.58 |
| Regulatory quality | 1990-1999 | 110 | 0.14 | 0.95 | -2.04 | 2.21 |
| Regulatory quality | 2000-2008 | 116 | 0.15 | 0.95 | -2.16 | 1.89 |
| Regulatory quality | All decades | 226 | 0.15 | 0.95 | -2.16 | 2.21 |
| Rule of law | 1990-1999 | 110 | 0.01 | 1.00 | -1.67 | 1.93 |
| Rule of law | 2000-2008 | 116 | 0.03 | 0.98 | -1.53 | 1.93 |
| Rule of law | All decades | 226 | 0.02 | 0.99 | -1.67 | 1.93 |
| Voice and accountability | 1990-1999 | 110 | 0.02 | 0.97 | -1.95 | 1.67 |
| Voice and accountability | 2000-2008 | 116 | 0.00 | 1.00 | -2.13 | 1.63 |
| Voice and accountability | All decades | 226 | 0.01 | 0.98 | -2.13 | 1.67 |
| Fitness Index | 1963-1969 | 102 | 0.99 | 2.59 | 0.00 | 15.5 |
| Fitness Index | 1970-1979 | 108 | 0.94 | 2.34 | 0.00 | 10.1 |
| Fitness Index | 1980-1989 | 103 | 1.00 | 4.54 | 0.00 | 45.6 |
| Fitness Index | 1990-1999 | 125 | 1.00 | 1.40 | 0.00 | 7.29 |
| Fitness Index | 2000-2008 | 128 | 1.00 | 1.21 | 0.00 | 5.81 |
| Fitness Index | All decades | 566 | 0.99 | 2.60 | 0.00 | 45.6 |
| Shannon Entropy | 1963-1969 | 102 | 2.76 | 1.26 | 0.04 | 5.26 |
| Shannon Entropy | 1970-1979 | 108 | 2.91 | 1.38 | 0.15 | 5.25 |
| Shannon Entropy | 1980-1989 | 103 | 3.22 | 1.50 | 0.33 | 5.65 |
| Shannon Entropy | 1990-1999 | 125 | 3.60 | 1.39 | 0.47 | 5.57 |
| Shannon Entropy | 2000-2008 | 128 | 3.50 | 1.39 | 0.27 | 5.51 |
| Shannon Entropy | All decades | 566 | 3.23 | 1.42 | 0.04 | 5.65 |
| Herfindahl-Hirschman | 1963-1969 | 102 | 0.23 | 0.23 | 0.01 | 0.99 |
| Herfindahl-Hirschman | 1970-1979 | 108 | 0.23 | 0.26 | 0.01 | 0.96 |
| Herfindahl-Hirschman | 1980-1989 | 103 | 0.21 | 0.24 | 0.01 | 0.91 |
| Herfindahl-Hirschman | 1990-1999 | 125 | 0.15 | 0.20 | 0.01 | 0.83 |
| Herfindahl-Hirschman | 2000-2008 | 128 | 0.17 | 0.21 | 0.01 | 0.91 |
| Herfindahl-Hirschman | All decades | 566 | 0.19 | 0.23 | 0.01 | 0.99 |

*Notes*: Descriptive summary statistics for all macro indicators and all decades intervals used in the paper and in the robustness checks presented in the SM. The time period, number of observations, mean, standard deviation, minimum and maximum are shown.

Table A2. *Clarke test for Gini EHII*

| Model 1 | Model 2 | Year | Clarke | Test-Stat |
|---|---|---|---|---|
| ECI | Fitness Index | Pooled | Model 1 is preferred (p < 2e-16) | 302 (81%) |
| ECI | Fitness Index | 2000-2008 | Model 1 is preferred (p = 9.1e-05) | 62 (71%) |
| ECI | Fitness Index | 1990-1999 | Model 1 is preferred (p = 2.5e-05) | 67 (72%) |
| ECI | Fitness Index | 1980-1989 | Model 1 is preferred (p = 1.0e-07) | 59 (81%) |
| ECI | Entropy | Pooled | Model 1 is preferred (p < 2e-16) | 287 (77%) |
| ECI | Entropy | 2000-2008 | Model 1 is preferred (p = 4.3e-06) | 65 (75%) |
| ECI | Entropy | 1990-1999 | Model 1 is preferred (p = 2.5e-05) | 67 (72%) |
| ECI | Entropy | 1980-1989 | Model 1 is preferred (p = 0.00037) | 52 (71%) |
| ECI | HHI | Pooled | Model 1 is preferred (p < 2e-16) | 296 (80%) |
| ECI | HHI | 2000-2008 | Model 1 is preferred (p = 4.3e-06) | 65 (75%) |
| ECI | HHI | 1990-1999 | Model 1 is preferred (p = 4.4e-10) | 76 (82%) |
| ECI | HHI | 1980-1989 | Model 1 is preferred (p = 4.1e-07) | 58 (79%) |
| ECI | Log GDP | Pooled | Model 1 is preferred (p = 2.8e-15) | 261 (70%) |
| ECI | Log GDP | 2000-2008 | Model 1 is preferred (p = 4.3e-07) | 67 (77%) |
| ECI | Log GDP | 1990-1999 | Model 1 is preferred (p = 3.5e-07) | 71 (76%) |
| ECI | Log GDP | 1980-1989 | Neither model is significantly preferred (p = 0.82) | 38 (52%) |

*Notes*: Clarke test compares Model 1 (ECI) with different measures of productive structure (Model 2). The dependent variable is GINI EHII.

Table A3. *Clarke test for Gini All*

| Model 1 | Model 2 | Year | Clarke | Test-Stat |
|---|---|---|---|---|
| ECI | Fitness Index | Pooled | Model 1 is preferred (p = 8.4e-14) | 248 (70%) |
| ECI | Fitness Index | 2000-2008 | Neither model is significantly preferred (p = 1) | 52 (50%) |
| ECI | Fitness Index | 1990-1999 | Model 1 is preferred (p = 0.0039) | 66 (65%) |
| ECI | Fitness Index | 1980-1989 | Model 1 is preferred (p = 2.8e-06) | 52 (79%) |
| ECI | Entropy | Pooled | Neither model is significantly preferred (p = 0.1) | 194 (54%) |
| ECI | Entropy | 2000-2008 | Model 1 is preferred (p = 0.019) | 65 (62%) |
| ECI | Entropy | 1990-1999 | Model 1 is preferred (p = 0.0039) | 66 (65%) |
| ECI | Entropy | 1980-1989 | Neither model is significantly preferred (p = 0.18) | 39 (59%) |
| ECI | HHI | Pooled | Model 1 is preferred (p = 7.8e-10) | 236 (66%) |
| ECI | HHI | 2000-2008 | Model 1 is preferred (p = 0.00039) | 71 (68%) |
| ECI | HHI | 1990-1999 | Model 1 is preferred (p = 3.9e-05) | 72 (71%) |
| ECI | HHI | 1980-1989 | Model 1 is preferred (p = 1.0e-04) | 49 (74%) |
| ECI | Log GDP | Pooled | Model 1 is preferred (p = 4.4e-13) | 246 (69%) |
| ECI | Log GDP | 2000-2008 | Model 1 is preferred (p = 0.00082) | 70 (67%) |

*Notes*: Clarke test compares Model 1 (ECI) with different measures of productive structure (Model 2). The dependent variable is GINI ALL.

# ROBUSTNESS CHECK FOR DIFFERENT INEQUALITY MEASURES

We find that ECI continues to be a negative and significant predictor of income inequality when the GINI ALL dataset is used instead of GINI EHII dataset (see Table A4). However, log GDP gains in significance as an inequality predictor. As was previously discussed, the GINI ALL dataset is strongly biased towards countries with a complex economy (see Figure A1), therefore is not a surprise that ECI loses some of its predictive power, since there is not much variation in ECI between countries in the sample.

Table A4. *Cross-Section Regression Results*

|  | \multicolumn{6}{c}{*Dependent variable: GINI ALL*} |  |  |  |  |  |
|---|---|---|---|---|---|---|
|  | (1) | (2) | (3) | (4) | (5) | (6) |
| ECI | -4.920*** |  | -3.640*** | -6.129*** | -3.967*** | -5.464*** |
|  | (1.166) |  | (1.213) | (1.172) | (1.091) | (1.200) |
| ln(GDP PPP pc) | 28.753*** | 27.002*** |  | 26.477*** | 27.364*** | 32.493*** |
|  | (4.978) | (5.221) |  | (5.160) | (4.994) | (4.623) |
| ln(GDP PPPpc)$^2$ | -1.648*** | -1.625*** |  | -1.580*** | -1.572*** | -1.776*** |
|  | (0.311) | (0.328) |  | (0.324) | (0.313) | (0.269) |
| Schooling | -1.257*** | -1.625*** | -0.850** |  | -1.362*** | -1.272*** |
|  | (0.325) | (0.330) | (0.340) |  | (0.325) | (0.331) |
| ln Population | 1.028** | 0.264 | 0.612 | 1.303*** |  | 0.832* |
|  | (0.477) | (0.464) | (0.522) | (0.492) |  | (0.435) |
| Rule of Law | -10.587*** | -10.759*** | -12.962*** | -12.343*** | -10.030*** |  |
|  | (2.509) | (2.640) | (2.715) | (2.576) | (2.525) |  |
| Corruption Control | 6.602*** | 8.087*** | 5.046* | 8.575*** | 5.882** |  |
|  | (2.472) | (2.575) | (2.664) | (2.524) | (2.477) |  |
| Government Effectiveness | -1.448 | -3.801 | -1.657 | -0.573 | -0.212 |  |
|  | (3.100) | (3.210) | (3.332) | (3.227) | (3.082) |  |
| Political Stability | -1.262 | -2.055* | -1.435 | -0.970 | -2.392** |  |
|  | (1.122) | (1.164) | (1.237) | (1.168) | (1.003) |  |
| Regulatory Quality | 5.971*** | 5.432** | 9.519*** | 5.152** | 4.941** |  |
|  | (2.231) | (2.345) | (2.381) | (2.319) | (2.205) |  |
| Voice and Accountability | 3.229** | 3.644*** | 3.260** | 2.344* | 3.240** |  |
|  | (1.324) | (1.390) | (1.448) | (1.361) | (1.339) |  |
| Constant | -89.788*** | -61.970*** | 34.531*** | -89.412*** | -65.991*** | -106.250*** |
|  | (22.344) | (22.471) | (9.433) | (23.323) | (19.651) | (21.661) |
| Observations | 167 | 167 | 167 | 167 | 167 | 167 |
| $R^2$ | 0.554 | 0.503 | 0.448 | 0.511 | 0.540 | 0.449 |
| Adjusted $R^2$ | 0.522 | 0.471 | 0.416 | 0.479 | 0.511 | 0.432 |
| Residual Std. Error | 6.632 (df = 155) | 6.980 (df = 156) | 7.332 (df = 157) | 6.922 (df = 156) | 6.709 (df = 156) | 7.229 (df = 161) |
| F Statistic | 17.490*** (df = 11; 155) | 15.759*** (df = 10; 156) | 14.135*** (df = 9; 157) | 16.284*** (df = 10; 156) | 18.346*** (df = 10; 156) | 26.273*** (df = 5; 161) |

*Note:* $^*$p<0.1; $^{**}$p<0.05; $^{***}$p<0.01

*Notes*: Pooled OLS regression models for GINI ALL. The regression table explores the effects of economic complexity on income inequality in comparison with other the socioeconomic factors, such as a country's average level of income, population, human capital and the institutional variables: corruption control, government effectiveness, political stability, regulatory quality, and voice and accountability. Column I includes all variables. Columns II-VI exclude blocks of variables to explore the contribution of each group of variables to the full model. The table pools data from two panels, one from 1996-2001 and another one from 2002-2008. If not otherwise indicated, the numbers in parenthesis are standard errors (SEM).

Next, we control the robustness of cross-sectional results within each decade. Table A5 shows that there is a negative and significant relationship between economic complexity and GINI EHII in all decades. Moreover, Table A6 presents the results of the same regression using GINI ALL as the dependent variable. In all decades, and in both tables, ECI results to be a negative and significant predictor on income inequality. The importance of GDP increases when using the GINI ALL dataset, but ECI continues to explain a significant percentage of the variance of income inequality.

Table A5. *Per decade cross-sections with GINI EHII*

| | *Dependent variable: GINI EHII* | | | | | | | | | |
|---|---|---|---|---|---|---|---|---|---|---|
| | 1963-69 | 1963-69 | 1970-79 | 1970-79 | 1980-89 | 1980-89 | 1990-99 | 1990-99 | 2000-08 | 2000-08 |
| ECI | -2.465*** (0.776) | | -1.819* (0.918) | | -2.374** (1.030) | | -4.193*** (0.864) | | -4.103*** (0.797) | |
| ln(GDP PPP pc) | 5.741 (5.213) | 8.231 (5.673) | 7.117 (5.734) | 9.977* (5.695) | 0.654 (4.840) | -0.036 (4.998) | -0.308 (3.310) | 1.371 (3.737) | 9.515** (3.886) | 8.842* (4.511) |
| ln(GDP PPPpc)$^2$ | -0.453 (0.345) | -0.701* (0.369) | -0.506 (0.376) | -0.737** (0.366) | -0.072 (0.291) | -0.094 (0.300) | 0.022 (0.200) | -0.185 (0.222) | -0.545** (0.220) | -0.603** (0.256) |
| Schooling | -0.743** (0.316) | -0.808** (0.347) | -0.909** (0.356) | -0.911** (0.366) | -0.815*** (0.304) | -1.020*** (0.301) | -0.636*** (0.235) | -1.175*** (0.235) | -0.631** (0.242) | -1.030*** (0.266) |
| ln Population | 0.191 (0.398) | 0.056 (0.436) | 0.019 (0.459) | -0.143 (0.463) | 0.089 (0.484) | -0.414 (0.447) | 0.533 (0.348) | -0.231 (0.352) | 0.700** (0.313) | 0.078 (0.335) |
| Constant | 26.432 (23.463) | 24.250 (25.815) | 21.614 (26.624) | 16.264 (27.181) | 44.467* (24.157) | 60.459** (23.941) | 40.546** (15.465) | 56.499*** (17.156) | -2.417 (18.085) | 20.036 (20.384) |
| Observations | 48 | 48 | 60 | 60 | 66 | 66 | 83 | 83 | 78 | 78 |
| R$^2$ | 0.822 | 0.779 | 0.676 | 0.652 | 0.569 | 0.531 | 0.715 | 0.627 | 0.683 | 0.567 |
| Adjusted R$^2$ | 0.800 | 0.758 | 0.646 | 0.627 | 0.533 | 0.500 | 0.696 | 0.608 | 0.661 | 0.543 |
| Residual Std. Error | 3.002 (df = 42) | 3.305 (df = 43) | 3.965 (df = 54) | 4.069 (df = 55) | 4.490 (df = 60) | 4.646 (df = 61) | 3.742 (df = 77) | 4.249 (df = 78) | 3.687 (df = 72) | 4.283 (df = 73) |
| F Statistic | 38.690*** (df = 5; 42) | 37.831*** (df = 4; 43) | 22.489*** (df = 5; 54) | 25.762*** (df = 4; 55) | 15.855*** (df = 5; 60) | 17.272*** (df = 4; 61) | 38.549*** (df = 5; 77) | 32.820*** (df = 4; 78) | 31.076*** (df = 5; 72) | 23.890*** (df = 4; 73) |

*Note:* $^*$p<0.1; $^{**}$p<0.05; $^{***}$p<0.01

*Notes*: Per decade cross-section regression with GINI EHII. The effect of ECI is negative and significant.

Table A6. *Per decade cross-sections with GINI All*

*Dependent variable: GINI All*

| | 1970-79 | 1970-79 | 1980-89 | 1980-89 | 1990-99 | 1990-99 | 2000-08 | 2000-08 |
|---|---|---|---|---|---|---|---|---|
| ECI | -2.298 | | -4.373** | | -6.294*** | | -3.989** | |
| | -1.824 | | -1.783 | | -1.867 | | -1.689 | |
| ln(GDP PPP pc) | 34.641*** | 38.609*** | 31.528*** | 33.530*** | 20.272*** | 23.484*** | 34.380*** | 34.429*** |
| | -9.487 | -9.013 | -8.794 | -9.153 | -6.589 | -6.911 | -6.829 | -7.013 |
| ln(GDP PPPpc)$^2$ | -2.122*** | -2.435*** | -1.866*** | -2.127*** | -1.037** | -1.405*** | -1.930*** | -2.036*** |
| | -0.626 | -0.579 | -0.55 | -0.564 | -0.405 | -0.414 | -0.397 | -0.405 |
| Schooling | -0.856 | -0.804 | -0.248 | -0.406 | -0.834* | -1.516*** | -1.096** | -1.427*** |
| | -0.562 | -0.564 | -0.54 | -0.561 | -0.484 | -0.466 | -0.474 | -0.464 |
| ln Population | 0.237 | 0.099 | 0.52 | -0.199 | 1.017 | 0.024 | 0.533 | 0.006 |
| | -0.767 | -0.765 | -0.768 | -0.743 | -0.665 | -0.632 | -0.611 | -0.584 |
| Constant | -95.121** | -104.858** | -95.314** | -82.141* | -65.237** | -46.523 | -108.258*** | -90.658*** |
| | -44.146 | -43.775 | -42.134 | -43.688 | -29.962 | -31.208 | -31.824 | -31.77 |
| Observations | 46 | 46 | 59 | 59 | 89 | 89 | 89 | 89 |
| $R^2$ | 0.542 | 0.524 | 0.449 | 0.387 | 0.359 | 0.271 | 0.403 | 0.363 |
| Adjusted $R^2$ | 0.485 | 0.478 | 0.397 | 0.341 | 0.32 | 0.237 | 0.367 | 0.332 |
| Residual Std. Error | 6.095 (df = 40) | 6.138 (df = 41) | 7.03 (df = 53) | 7.35 (df = 54) | 7.756 (df = 83) | 8.220 (df = 84) | 7.506 (df = 83) | 7.708 (df = 84) |
| F Statistic | 9.481*** (df = 5; 40) | 11.293*** (df = 4; 41) | 8.643*** (df = 5; 53) | 8.509*** (df = 4; 54) | 9.301*** (df = 5; 83) | 7.819*** (df = 4; 84) | 11.200*** (df = 5; 83) | 11.955*** (df = 4; 84) |

*Note:* *p<0.1; **p<0.05; ***p<0.01

*Notes*: Per decade cross-section regression with Gini All. The effect of ECI is negative and significant.

# EFFECTS OF OTHER MEASURES OF EXPORT DIVERSITY, COMPLEXITY AND CONCENTRATION

Table A7 reproduces the cross-sectional regression Table 1 from the paper, with the addition of the Fitness Index, Shannon Entropy, and the Herfindahl-Hirschman Index (HHI). The results show that all of these measures are significant when included in the regression individually, however ECI explains a larger fraction of the variance in inequality. Notably, a higher economic concentration seems to lead to a higher level of income inequality, however this effect is not significant if economic complexity (ECI) is included.

Table A7. *Cross-Section Regression Results*

| | \multicolumn{6}{c}{Dependent variable: GINI EHII} | | | | | |
|---|---|---|---|---|---|---|
| | (1) | (2) | (3) | (4) | (5) | (6) |
| ECI | -0.040*** | | | | | -0.036*** |
| | (0.007) | | | | | (0.007) |
| Fitness Index | | -0.023*** | | | | |
| | | (0.005) | | | | |
| Entropy | | | -0.025*** | | | |
| | | | (0.005) | | | |
| HHI | | | | 0.146*** | | 0.058 |
| | | | | (0.044) | | (0.044) |
| ln(GDP PPP pc) | 0.067** | 0.036 | 0.086*** | 0.065** | 0.059* | 0.068** |
| | (0.028) | (0.029) | (0.029) | (0.031) | (0.032) | (0.028) |
| ln(GDP PPPpc)$^2$ | -0.004** | -0.002 | -0.005*** | -0.004** | -0.004* | -0.004** |
| | (0.002) | (0.002) | (0.002) | (0.002) | (0.002) | (0.002) |
| Schooling | -0.005*** | -0.008*** | -0.008*** | -0.008*** | -0.009*** | -0.005*** |
| | (0.002) | (0.002) | (0.002) | (0.002) | (0.002) | (0.002) |
| ln Population | 0.007** | 0.010*** | 0.008*** | 0.004 | 0.0001 | 0.007*** |
| | (0.003) | (0.003) | (0.003) | (0.003) | (0.003) | (0.003) |
| Rule of Law | -0.013 | -0.008 | -0.013 | -0.017 | -0.016 | -0.014 |
| | (0.013) | (0.013) | (0.013) | (0.014) | (0.014) | (0.013) |
| Corruption Control | 0.011 | 0.009 | 0.011 | 0.019 | 0.027* | 0.009 |
| | (0.013) | (0.014) | (0.013) | (0.014) | (0.014) | (0.013) |
| Government Effectiveness | 0.002 | -0.013 | -0.007 | -0.012 | -0.022 | 0.003 |
| | (0.017) | (0.017) | (0.017) | (0.018) | (0.018) | (0.017) |
| Political Stability | -0.010 | -0.011* | -0.014** | -0.017** | -0.017** | -0.011* |
| | (0.006) | (0.007) | (0.006) | (0.007) | (0.007) | (0.006) |
| Regulatory Quality | -0.006 | -0.006 | 0.001 | -0.0002 | -0.012 | -0.002 |
| | (0.012) | (0.013) | (0.013) | (0.014) | (0.014) | (0.013) |
| Voice and Accountability | 0.001 | 0.009 | 0.015* | 0.011 | 0.006 | 0.004 |
| | (0.008) | (0.008) | (0.008) | (0.008) | (0.009) | (0.008) |
| Constant | 0.083 | 0.199 | 0.132 | 0.206 | 0.286** | 0.071 |
| | (0.130) | (0.131) | (0.132) | (0.138) | (0.141) | (0.130) |
| Observations | 142 | 142 | 142 | 142 | 142 | 142 |
| R$^2$ | 0.717 | 0.698 | 0.703 | 0.667 | 0.639 | 0.721 |
| Adjusted R$^2$ | 0.693 | 0.672 | 0.678 | 0.639 | 0.612 | 0.695 |
| Residual Std. Error | 0.035 (df = 130) | 0.036 (df = 130) | 0.035 (df = 130) | 0.037 (df = 130) | 0.039 (df = 131) | 0.034 (df = 129) |
| F Statistic | 29.916*** (df = 11; 130) | 27.282*** (df = 11; 130) | 28.014*** (df = 11; 130) | 23.698*** (df = 11; 130) | 23.208*** (df = 10; 131) | 27.720*** (df = 12; 129) |

*Note:* *p<0.1; **p<0.05; ***p<0.01

*Notes*: Pooled cross-section regression using different measures of productive structure, using GINI EHII as a dependent variable. ECI, Fitness Index, and Entropy are negatively correlated with Gini, and HHI is positively correlated with Gini. All regression coefficients between Gini and the four measures of productive structure are significant.

Next, we compare the effects of ECI (Hidalgo and Hausmann, 2009), the Fitness Index (Tacchella et al., 2012), Shannon-Entropy (Shannon, 1948) and the Herfindahl-Hirschman Index (Hirschman, 1945, Herfindahl, 1950) in fixed-effects panel regression (Table A8). The results show that, among these measures of productive structure, ECI is the only measure which is a significant predictor of time variations in inequality within countries over long periods of time.

Table A8. *Fixed-Effects Regression Results GINI EHII*

| | *Dependent variable: GINI EHII* | | | | | |
|---|---|---|---|---|---|---|
| | (1) | (2) | (3) | (4) | (5) | (6) |
| ECI | -2.189*** <br> (0.555) | | | | | -2.302*** <br> (0.570) |
| Fitness Index | | -0.070 <br> (0.075) | | | | |
| Entropy | | | -0.067 <br> (0.408) | | | |
| HHI | | | | 0.080 <br> (2.352) | | -2.054 <br> (2.340) |
| ln(GDP PPP pc) | -1.541 <br> (2.915) | -2.749 <br> (3.009) | -3.285 <br> (3.044) | -3.171 <br> (2.996) | -3.182 <br> (2.973) | -1.735 <br> (2.925) |
| ln(GDP PPPpc)$^2$ | -0.019 <br> (0.181) | 0.015 <br> (0.188) | 0.049 <br> (0.189) | 0.043 <br> (0.187) | 0.044 <br> (0.186) | -0.002 <br> (0.182) |
| Schooling | 1.380*** <br> (0.282) | 1.475*** <br> (0.291) | 1.444*** <br> (0.295) | 1.454*** <br> (0.294) | 1.453*** <br> (0.290) | 1.334*** <br> (0.287) |
| ln Population | -2.213** <br> (1.098) | -1.724 <br> (1.124) | -1.708 <br> (1.142) | -1.671 <br> (1.133) | -1.675 <br> (1.122) | -2.357** <br> (1.111) |
| Observations | 335 | 335 | 335 | 335 | 335 | 335 |
| R$^2$ | 0.216 | 0.167 | 0.164 | 0.164 | 0.164 | 0.219 |
| Adjusted R$^2$ | 0.152 | 0.117 | 0.115 | 0.115 | 0.116 | 0.153 |
| Residual Std. Error | 12.959*** <br> (df = 5; 235) | 9.448*** <br> (df = 5; 235) | 9.242*** <br> (df = 5; 235) | 9.236*** <br> (df = 5; 235) | 11.594*** <br> (df = 4; 236) | 10.917*** <br> (df = 6; 234) |

*Note:* $^*$p<0.1; $^{**}$p<0.05; $^{***}$p<0.01

*Notes*: Effects of different diversity measures in fixed-effects panel regressions, using GINI EHII

# PRODUCT GINI INDEX

This section analyzes the evolution of PGIs over time, and to which extent complex products are also more inclusive products.

**PGI ranking**

Table A9 shows the highest and lowest ranked five products in the PGI index between 1995-2008. A full list of all 775 products can be found in a separate Appendix Table A12.

Table A9. *List of the 5 products with the highest and lowest PGI values between 1995-2008*

**5 products with highest PGI**

| SITC4 | Product Name | Product Section | PGI |
|---|---|---|---|
| 721 | Cocoa Beans | Food and live animals | 0.506 |
| 814 | Inedible Flours of Meat and Fish | Food and live animals | 0.505 |
| 2683 | Fine Animal Hair | Crude materials, inedible, except fuels | 0.503 |
| 6545 | Jute Woven Fabrics | Manufactured goods classified chiefly by material | 0.499 |
| 2875 | Zinc Ore | Crude materials, inedible, except fuels | 0.498 |

**5 products with lowest PGI**

| SITC4 | Product Name | Product Section | PGI |
|---|---|---|---|
| 7259 | Paper Making Machine Parts | Machinery and transport equipment | 0.334 |
| 7244 | Textile Machinery | Machinery and transport equipment | 0.336 |
| 7233 | Road Rollers | Machinery and transport equipment | 0.338 |
| 2120 | Raw Furs | Crude materials, inedible, except fuels | 0.338 |
| 7252 | Paper Making Machines | Machinery and transport equipment | 0.340 |

*Notes*: PGI Ranking: List of the 5 products with the respective highest and lowest PGI values between 1995-2008.

**Descriptive statistics and evolution of PGIs**

As discussed in the paper, the PGIs associate each product with a level of income inequality by calculating the average Gini coefficient of the countries that produce the respective product, weighted by the product's importance in the country's economy. Table A10 shows summary statistics for the PGIs for each decade intervals. Due to limited data availability we exclude the period between 1963 and 1969.

The average value of the PGIs increases over time, representing the trend found in recent research on inequality measures that countries are converging towards an average Gini value around 0.40 (Palma, 2011). However, despite this convergence trend, the spread between minimum and maximum value of the PGIs remains large. While the PGI value varied between 0.285 to 0.511 in the time period between 1970-1979, in the time period between 2000-2008 the values were distributed between 0.334 and 0.517.

Table A11 illustrates that the PGI values for different decades are highly correlated with each other, and that this correlation—as expected—tends to decline over time.

Table A10. *Descriptive statistics of PGI values for different decades*

| Time period | Mean | Std. | Min | Mmax |
|---|---|---|---|---|
| **1970-1979** | 0.367 | 0.043 | 0.285 | 0.512 |
| **1980-1989** | 0.383 | 0.042 | 0.305 | 0.504 |
| **1990-1999** | 0.403 | 0.039 | 0.327 | 0.496 |
| **2000-2008** | 0.418 | 0.038 | 0.337 | 0.518 |

Table A11. *Correlation coefficient between PGI values from different decades*

| PGI value | 1970-79 | 1980-89 | 1990-99 | 2000-08 |
|---|---|---|---|---|
| **1970-1979** | 1.000 | 0.879 | 0.734 | 0.667 |
| **1980-1989** | | 1.000 | 0.841 | 0.746 |
| **1990-1999** | | | 1.000 | 0.869 |
| **2000-2008** | | | | 1.000 |

**Correlation between PGI and product complexity**

Figure A2 illustrates a strong and negative correlation between PGI and the Product Complexity Index (PCI) (Hidalgo & Hausmann, 2009; Hausmann et al., 2014; Felipe et al., 2012) for different decades. In other words, more complex industrial products tend to be associated with lower levels of inequality.

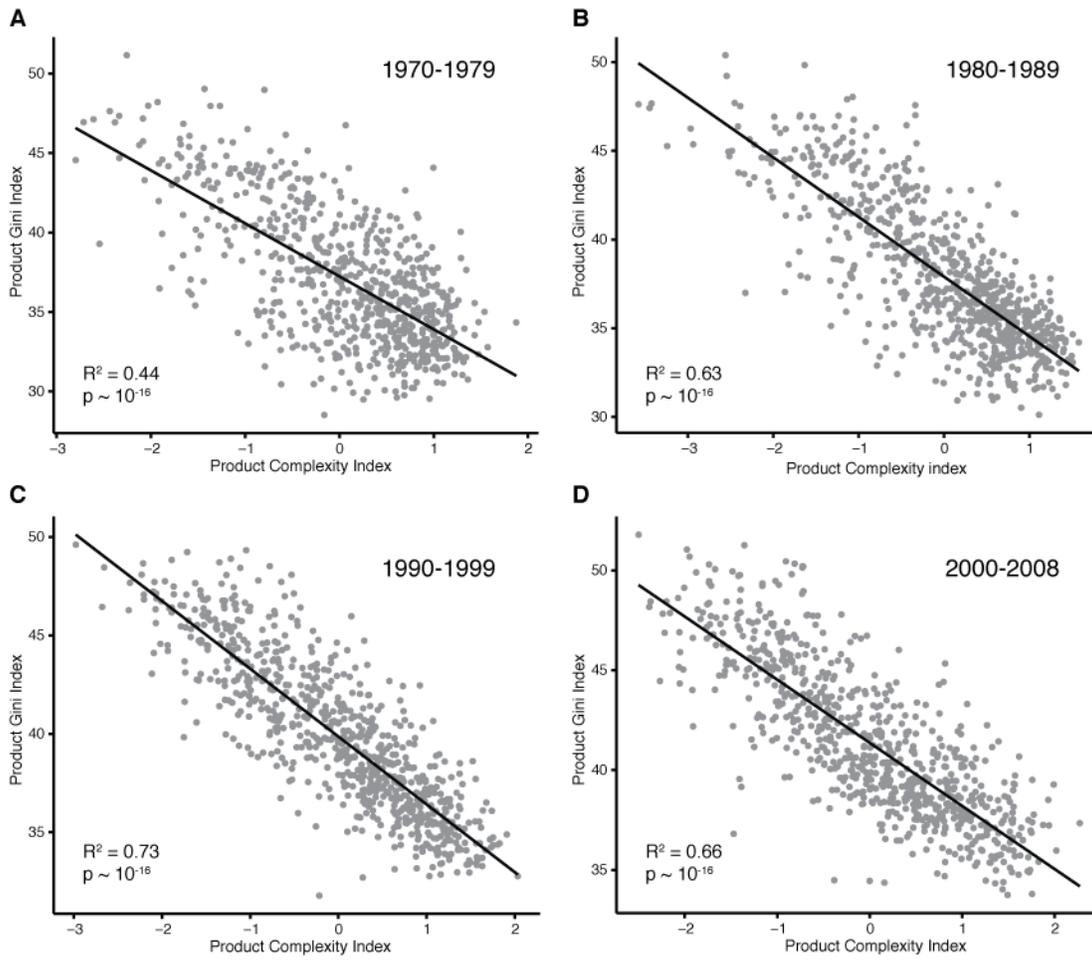

Figure A2. *Bivariate relationship between the Product Complexity Index (PCI) and the Product Gini Index (PGI) in the (A) 1970-1979, (B) 1980-1989, (C) 1990-1999 and (D) 2000-2008*